**A novel approach for modeling the non-Newtonian behavior of simple liquids: application to liquid water viscosity from low to high shear rates**

F. Aitken, F. Volino

Univ. Grenoble Alpes, CNRS, Grenoble INP, G2ELab, F-38000 Grenoble, France.

**Abstract**. The aim of this paper is to present a modeling for the rheological behavior of simple liquids as a function of the amplitude of the imposed shear stress or strain. The elastic mode theory (Ref. 6) is first generalized to take into account the fact that during a flow experiment, mechanical energy is injected in a system initially at thermodynamic equilibrium. This generalized theory can be seen as a particular aspect of the general problem of perturbation by the measurement, associated with that of the coupling between fluctuation and dissipation. This generalization leads to a "finitary" character of the model. It is then combined with the inertial mode theory (Ref. 7). The formalism thus obtained allows to model the rheological behavior of liquids over a wide range of velocity gradients, including the intermediate narrow range corresponding to the Newtonian regime. As experimental tests, viscosity measurements with two kinds of moving rotor rheometers were performed. Only data obtained with liquid water at room temperature are presented and quantitatively analyzed here. It is also shown that liquid n-octane exhibits the same qualitative behaviors as those of liquid water. In the appendices, connection of this theory with quantum mechanics and turbulence phenomena are discussed, and the notion of viscous mass is introduced.

## 1 Introduction

The knowledge of rheological properties of fluids is fundamental for understanding their physical structure and for the development of a large number of technical and industrial applications. Among the recent applications where these properties are useful for their optimization, we can mention for example the development of microelectromechanical systems (MEMS) using liquids (e.g. see Ref. 1) and especially those based on adiabatic logic (e.g. see Ref. 2).

The rheological properties of fluid media are generally interpreted through the concept of dynamic viscosity. The viscosity is defined by Newton's postulate such that the shear stress $\sigma$ is proportional to the velocity gradient $\nabla v_f$, or shear rate, where the proportionality factor is by definition the viscosity $\eta$, i.e. $\sigma = \eta \nabla v_f$. A particular class of behavior is the so-called Newtonian fluid where the viscosity is independent of the velocity gradient. But experiments show that for all fluid media, the viscosity is a function of the velocity gradient except for a more or less extended range called the Newtonian plateau. For the vast majority of fluid media, at low velocity gradients (i.e. lower than those corresponding to the Newtonian plateau), the viscosity decreases when the velocity gradient increases (e.g. see chap. 2 of Ref. 3). This behavior is called "shear-thinning". On the other hand, when the values become higher than those defining the Newtonian plateau, the viscosity generally increases with the increase of the velocity gradient (e.g. see chap. 2 of Ref. 3). This behavior is called "shear-thickening" (or "dilatancy"). The interest in understanding these different rheological behaviors was written by Heyes (Ref. 4) in the following terms:

> "we believe that all liquids exhibit similar non-Newtonian behaviour [...] and therefore it has a fundamental physical and theoretical significance."

These deviations from the Newtonian behavior for simple liquids such as liquid argon, nitrogen, chlorine and water are generally interpreted from a microscopic point of view in terms of molecular rearrangements and structuring of the medium on large scales in front of the molecular scale but the mechanisms in terms of molecular dynamics are poorly understood and unsatisfactory (e.g. Refs. 4 and 5).

In this paper, we will analyze these different viscosity behaviors as a function of the velocity gradient using the elastic mode theory developed in Ref. 6 in association with the inertial mode theory developed in Ref. 7. But to do so, the elastic mode theory of Ref. 6 must be generalized when mechanical energy introduced in the system to produce the flow is no longer negligible in front of the shear elastic energy associated with the medium at thermodynamic equilibrium.

We then apply this modeling to analyze viscosity data of liquid water for which all the parameters of the theory have been determined in Refs. 6 and 7. It will be shown that this generalized theory is numerically equivalent to the restricted theory developed in Ref. 6 for analyzing the usual viscometry experiments corresponding to the Newtonian plateau. In other words, in the vast majority of usual viscosity measurements on simple liquids, it can be said that the influence of the perturbation by the measurement can be neglected. We then analyze the data we obtained using two kinds of experiments with moving rotor rheometers where the velocity gradient can be varied in a broad range including the Newtonian plateau. An important advantage of these devices is that the velocity gradients can be easily determined in the whole volume, contrarily to the usual viscometry devices. The experiments with liquid water at atmospheric pressure and room temperature successively show shear-thinning, Newtonian plateau and then shear-thickening behavior. We then show that the generalized theory allows to account quantitatively for these various behaviors in a consistent way (e.g. see Fig. 14d). The interpretation that emerges is that the shear-thinning behavior is the consequence of the solid-liquid dynamic phase transition of the liquid associated with a variation of the correlation length while the shear-thickening behavior is the consequence of the increase of the influence of the mechanical energy injected in the system which leads to the reduction of the amplitude of the thermal fluctuations. Classical mechanics is recovered for very large gradients since in this case the effective amplitude of the thermal fluctuations tends to zero.

In the appendices, we explore three consequences of the theory. In the first one, a theoretical expression whose form is reminiscent of the uncertainty principle of quantum mechanics is discussed. In the second one, a general criterion for the transition to turbulence is given using a number whose value is equal to unity at the transition unlike the Reynolds number. In the third one, we introduce and discuss the notion of *viscous mass* which is a fundamental quantity of the theory.

## 2 Influence of external actions in the elastic mode theory

When a system is at thermodynamic equilibrium, the excess elastic energy is due to random displacements of objects from their equilibrium position. Now when a system is put out-of-equilibrium by an external action, the energy introduced into the system by this action must be added to the excess elastic energy. So in this section, we will first describe in a general way how this external energy must be taken into account in the elastic energy functional and then we will particularize the modeling in the case where the external energy is associated with translational macroscopic motions corresponding to fluid flows.

### 2.1. Generalized expression of the elastic energy functional

Elastic mode theory has been described in detail in Ref. 6, so we refer the reader to this article for more details. In this section, only those elements that are modified by the introduction of external actions will be described.

Let $\vec{a}(\vec{r})$ be a random variable of components $(a_x, a_y, a_z)$ which describes the fluctuation on a lattice of a basic unit (i.e. a set of $n_B$ molecules or atoms) property due to the non-zero temperature of the system at equilibrium. The random variable $\vec{a}$ can optionally represent the instantaneous position $\vec{u}$ or the orientation $\vec{\Omega}$ of the local director of a basic unit. The starting point of this model is the assumption that $\vec{a}(\vec{r})$ can be developed into Fourier series (whose coefficients refer to as *elastic modes*) on the lattice. So, for component $a_x$ of $\vec{a}$, one has:

$$a_x(\vec{r}) = \sum_{\vec{q}} a_x(\vec{q}) e^{i\vec{q}\cdot\vec{r}} = \sum_{\vec{q}} a_{xq} e^{i\vec{q}\cdot\vec{r}} \qquad (1)$$

where the amplitudes $a_{xq}$ are new statistically independent random variables. Each mode is characterized by its wave-vector $\vec{q}$ and its polarization.

As in Ref. 6, the isotropy of reciprocal space is always considered in such a way that the wave-vector moduli $q$ are limited at short length scales by a cut-off wave-vector $q_c$, and towards long length scales by a wave-vector $q_c / N$, where $N/q_c$ represents the fluctuative distance (i.e. the coherence length) compatible with the sample size.

From the energetic point of view, the fluctuation of $\vec{a}$ around its equilibrium position is to increase the energy of the sample compared to the perfectly ordered state at zero Kelvin. As in Ref. 6, this excess energy due to thermal fluctuations can be described by an elastic energy functional $F_{eq}$ such that:

$$F_{eq} = \frac{1}{2} \int K_a \left[ \vec{\nabla}^\alpha \vec{a} \right]^2 dV \qquad (2)$$

where $\vec{\nabla}^\alpha$ stands for the fractional gradient operator (that is the derivative of order $\alpha$ with respect to $\vec{r}$), and where the coefficient $K_a$ is an elastic constant.

Now, when an external action puts the system out-of-equilibrium, then it is postulated that the functional elastic energy is written:

$$F = F_{eq} + E_{ext} \qquad (3)$$

where $F_{eq}$ is given by Eq. (2) and $E_{ext} = \frac{1}{2}\int f_E dV$, $f_E$ being the energy per unit volume which describes the coupling between random microscopic motions and deterministic macroscopic motions.

In all that follows, it will be assumed that the external energy term can always be written in the following form:

$$E_{ext} = \frac{1}{2}\bar{f}_E^a \int \bar{a}^2 dV \qquad (4)$$

where $\bar{f}_E^a$ represents an average value associated with the external energy per unit volume but whose dimensionality depends on the nature of $\bar{a}$.

In addition, the fractional exponent $\alpha$ will be considered in the following to be such that $\alpha = 1 + v/2$. By developing the components of $\bar{a}$ in spatial Fourier series and by using the orthogonality of complex exponential functions, it is deduced that the functional $F$ is written in the following quadratic form:

$$F = \frac{1}{2}V\sum_q \left( K_a q^2 \left(\frac{q}{q_c}\right)^v + \bar{f}_E^a \right) \left| a_q^2 \right| \qquad (5)$$

where $\left| a_q^2 \right| = \left| a_{kq}^2 \right|, \forall k = x, y, z$ corresponding to an isotropic property. By combining Eq. (5) with the assumption of the equipartition of thermal energy, namely that the average energy per mode is $k_B T/2$, where $k_B$ is the Boltzmann constant and $T$ the absolute temperature, and integrating over all $q$ modes, the following result is obtained for the expression of the fluctuations full mean square $< \left| a^2 \right| >$ due to transverse modes only:

$$< \left| a^2 \right| > = \frac{k_B T q_c}{\pi^2 K_a} H_N\left(v, \bar{\lambda}\right) \qquad (6)$$

with

$$H_N\left(v, \bar{\lambda}\right) = \int_{1/N}^{1} \frac{x^2 dx}{x^{2+v} + \bar{\lambda}} \qquad (7)$$

and where $\bar{\lambda} = \bar{f}_E^a / \left( K_a q_c^2 \right)$ is a non-dimensional quantity. Eq. (7) shows that as soon as the parameter $\bar{\lambda}$ is non-zero (i.e. as soon as an action disturbs the system), the denominator of the integrand never becomes zero and therefore the integral converges systematically whatever the value of $N$ (finite or infinite). Since, in real experiments, the studied system is, strictly speaking, always out-of-equilibrium because it is necessary to exchange energy with it to obtain any information, it follows that the present modeling implies that "infinities" do not exist in experimental physics. In this sense, we can say that this model is "finitary".

When $\bar{\lambda} = 0$, the function $H_N(v)$ introduced in Ref. 6 is recovered such that:

$$H_N\left(v,\bar{\lambda}=0\right)=H_N(v)=\frac{N^{v-1}-1}{v-1} \qquad (8)$$

Now, it is observed that any value of $\bar{\lambda}>0$ implies that $H_N\left(v,\bar{\lambda}\right)$ will have a smaller value than $H_N(v)$ for the same value of $v$ and $N$. The immediate consequence is that any external action leads to a reduction in the fluctuations of $<\left|a^2\right|>$ compared to what they would be in the absence of external action. In the extreme limit where $\bar{\lambda}\to\infty$, it is deduced by making a series development of the integrand in Eq. (7) that:

$$H_N\left(v>-2,\bar{\lambda}\to\infty\right)\approx\frac{1}{3\bar{\lambda}}\left(1-\frac{1}{N^3}\right)\xrightarrow[\forall N]{}0 \qquad (9)$$

The approximated expression in Eq. (9) shows when $\bar{\lambda}>>1$ and $N$ tends towards infinity, $H_N\left(v,\bar{\lambda}\right)$ tends towards a finite value $\left(3\bar{\lambda}\right)^{-1}$ whereas previously with Eq. (8) the limit was such that $\lim_{N\to\infty}H_N(v>1)\to\infty$.

It is instructive to study the behavior of the function $H_N\left(v,\bar{\lambda}\right)$ in intermediate cases where $\bar{\lambda}$ is neither null nor infinitely large. By construction, $\bar{\lambda}$ is independent of $q$ and therefore it can be deduced an analytical expression of Eq. (7) such as:

$$H_N\left(v>-2,\bar{\lambda}\right)=\frac{N^{v-1}{}_2F_1\left(1,\frac{v-1}{v+2};\frac{2v+1}{v+2};-\bar{\lambda}N^{2+v}\right)-{}_2F_1\left(1,\frac{v-1}{v+2};\frac{2v+1}{v+2};-\bar{\lambda}\right)}{v-1} \qquad (10)$$

where ${}_2F_1\left(a,b;c;z\right)=\sum_{k=0}^{\infty}\frac{(a)_k(b)_k}{(c)_k}\frac{z^k}{k!}$ is the hypergeometric function such that $(\bullet)_k$ represents the Pochhammer symbol. Despite the appearance of the hypergeometric functions, a great similarity between Eq. (8) and Eq. (10) can be noticed. It is easy to check that the hypergeometric functions involved in Eq. (10) are such that they tend towards the value of 1 for all $v>-2$ when $\bar{\lambda}$ tends towards 0, in other words, Eq. (8) is recovered at the limit $\bar{\lambda}=0$. Given that $N$ is generally a large number in front of 1, it is deduced from Eq. (10) that $H_N\left(v,\bar{\lambda}\right)$ will be little different from $H_N(v)$ (i.e. $H_N\left(v,\bar{\lambda}\right)\cong H_N(v)$) when:

$$\bar{\lambda}<<\frac{1}{N^{2+v}} \qquad (11)$$

In the particular case where $\bar{\lambda}$ is a parameter independent of $v$, Fig. 1 shows the evolution of Eq. (8) and Eq. (10) for a fixed value of $v$. Different values of $v$ do not qualitatively change in this case the shape of the surfaces. It can be observed that the functions $H_N\left(v,\bar{\lambda}\right)$ and $H_N(v)$ are almost indistinguishable regardless of the value of $\bar{\lambda}$ as soon as $N$ is less than 10. Now when $N$ increases, it follows that the approximation $H_N\left(v,\bar{\lambda}\right)\cong H_N(v)$ is obtained for lower and lower values of $\bar{\lambda}$ as indicated by Eq. (11). On the other hand, for $\bar{\lambda}>>1$, $H_N\left(v,\bar{\lambda}\right)$ tends towards zero whatever the value of $N$ according to Eq. (9).

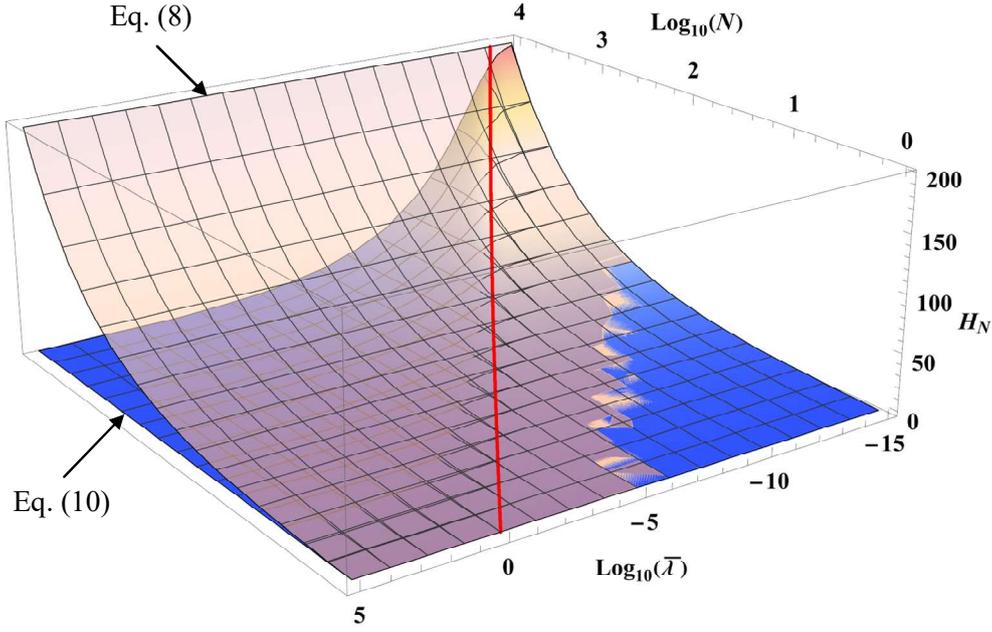

Fig. 1. Semi-logarithmic plot of the variations of Eq. (8) and Eq. (10) when $\overline{\lambda}$ is a constant independent of $v$ for $v = 1.5$. The thick red curve represents the limit of validity of $H_N(v)$ when $(\overline{\lambda})_{\text{threshold}} = N^{-(2+v)}$ as given by Eq. (11).

In the particular case of Fig. 1, Eq. (9) shows that for $\overline{\lambda}$ sufficiently large $H_N\left(v, \overline{\lambda}\right)$ becomes independent of $N$ and $v$, which means that the thermal phase transition aspect is completely deleted by the "intensity" of the external action. This can also be understood as follows: the effect of the external action $\overline{\lambda}$ can be assigned to the quantity $N$ in such a way that from Eq. (8) and Eq. (10) it is possible to define an effective $N$ that can be written:

$$N_{\text{eff}} = \left(1 + N^{v-1} {}_2F_1\left(1, \frac{v-1}{v+2}; \frac{2v+1}{v+2}; -\overline{\lambda}N^{2+v}\right) - {}_2F_1\left(1, \frac{v-1}{v+2}; \frac{2v+1}{v+2}; -\overline{\lambda}\right)\right)^{\frac{1}{v-1}} \quad (12)$$

Fig. 2 shows that when $\overline{\lambda}$ is large enough, whatever the value of $N$, one finds a $N_{\text{eff}}$ that tends towards the value 1, and it is then clear that the very notion of phase transition for a sample formed of independent particles makes no sense. Moreover, Eq. (8) implies that $H_{N_{\text{eff}}}(v)$ tends to zero, so $<\left|a^2\right|>$ tends also to zero, when $N_{\text{eff}}$ tends to 1. In other words, when the particles have been made independent of each other by a sufficiently intense external action, the size of the thermal cloud is zero and the particle behaves like a classical object of the material point mechanics.

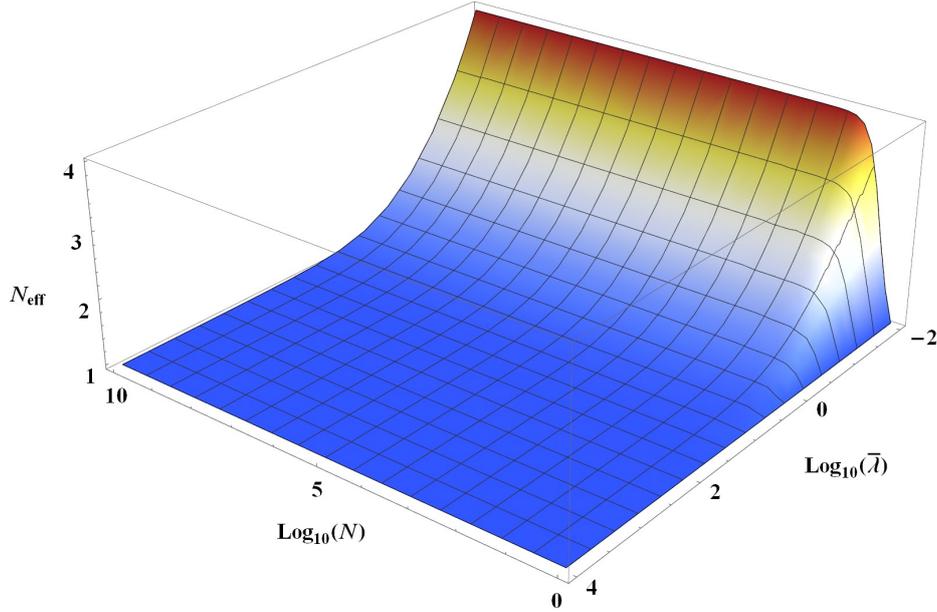

Fig. 2. Semi-logarithmic plot of the variations of Eq. (12) when $\overline{\lambda}$ is a constant independent of $v$ for $v = 1.5$.

The previous analytical relationships are valid as long as $v > -2$. In the case where $v$ is smaller than -2 then the calculation of $H_N\left(v, \overline{\lambda}\right)$ defined by Eq. (7) must be done in numerical form.

## 2.2. Application to translational macroscopic motion

In this section, the developments of section 2.1 are applied to the case of fluid flows, i.e. when the external energy is associated with translational macroscopic motions. Thus, in the above equations, the vector $\vec{a}$ must be identified with the displacement vector $\vec{u}$ and the constant $K_a$ with the shear elastic constant $K$ defined in Ref. 6. The fluid phases correspond to the disordered phase where the exponent $v$ is given by the relation $v - 1 = \left(1 - \dfrac{T_t}{T}\right)^{\frac{1}{4}}$, $T_t$ being a temperature associated with the glass transition that occurs when $T_t = T$. In this case, it is assumed that $\overline{\lambda}$ is written in the following form:

$$\overline{\lambda} = \lambda^{1+v}; \ \ v \geq 1 \tag{13}$$

where $\lambda$ is a new parameter independent of $v$. Equation (13) implies that the importance of the external action depends on the distance at which the phase transition is located: for $\lambda < 1$, the reduction of fluctuations on $<\left|u^2\right|>$ is a decreasing value of $v$ and for $\lambda >> 1$, $<\left|u^2\right|>$ becomes practically independent of $v$. Fig. 3 shows that this allows greater latitude on the values of parameter $\lambda$ for which one has $H_N\left(v, \lambda\right) \approx H_N\left(v\right)$ in accordance with Eq. (11) now being written:

$$\lambda << \dfrac{1}{N^{\frac{2+v}{1+v}}}; \ \ v \geq 1 \tag{14}$$

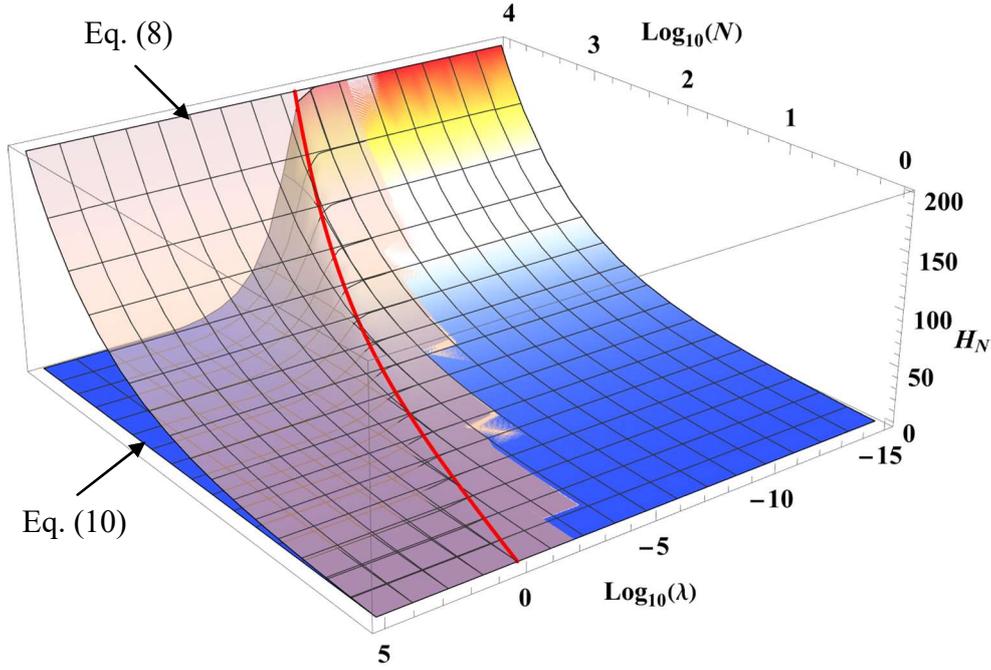

Fig. 3. Semi-logarithmic plot of the variations of Eq. (8) and Eq. (10) when $\bar{\lambda}$ is given by Eq. (13) for $v = 1.5$. The thick red curve represents the limit of validity of $H_N(v)$ when $(\lambda)_{\text{threshold}} = N^{-\frac{2+v}{1+v}}$ as given by Eq. (14).

Equation (13) and the expression of the parameter $\lambda$ will be justified in section 3 when analyzing the experimental data. Indeed, the parameter $\lambda$ is related to the external action (i.e. it is zero for a system at equilibrium). However, the macroscopic motions associated with the flow are characterized by a linear velocity gradient field related to the rate of viscous dissipation. From Eq. (4), it is expected that $\lambda$ is a function of the average velocity gradient $(\nabla v_f)_{\text{av}}$, where $v_f$ represents the velocity of a fluid particle as defined in Ref. 7. The parameter $\lambda$ being a non-dimensional number just like the Reynolds number Re, it is then interesting as an example to express Re in term of the parameters of the model with the approximation $(v_f)_{\text{av}} \cong (\nabla v_f)_{\text{av}} d$ where $d$ is the dissipative distance. Moreover, in liquid phase, according to Ref. 6, the following approximation of the fluid dynamic viscosity $\eta \approx \eta_l$ can be made. Let's remember that within the limit where Eq. (14) is satisfied, we have $\eta_l = \dfrac{K}{H_N(v)} \tau$ with $\tau = d/c_0$, where $c_0 = \sqrt{K/\rho}$ is the characteristic shear elastic celerity in the medium. Then, it comes:

$$\text{Re} = \frac{\rho (v_f)_{\text{av}} L}{\eta} \cong \frac{L}{d} (\nabla v_f)_{\text{av}} \tau \, H_N(v) \tag{15}$$

where $L$ represents the hydraulic diameter corresponding to the particular experiment considered and $\tau \, H_N(v)$ the time scale of viscous diffusion when the Knudsen term of the viscosity can be neglected. We recall that the more general expression is given by $\tau_v = \dfrac{d^2}{\eta/\rho}$ (Ref. 7) which represents the ratio between the kinetic energy and the average power dissipated, per unit volume.

For example, if we consider the flow in a tube, then $L$ corresponds to the tube diameter $D$ and if the diameter is sufficiently small then the dissipative distance is such that $d = D/2$. From Eq. (15), one deduces that:

$$\mathrm{Re} \cong 2\left(\nabla v_f\right)_{\mathrm{av}} \tau\, H_N(\nu) \qquad (16)$$

This Eq. (16) shows that the Reynolds number is scaled by the product $\Lambda = \left(\nabla v_f\right)_{\mathrm{av}} \tau$. The quantity $\left(\nabla v_f\right)_{\mathrm{av}}$ represents an average rate of injection of mechanical energy by the external source, while $\tau$ represents the average time required by the system to evacuate this energy to the outside. This product $\Lambda$, which involves the macroscopic distance $d = c_0\,\tau$, is therefore associated with a global property of the flowing system.

Given that by construction $\lambda$ is zero for a system at equilibrium, it is logical to assume that $\lambda$ is an increasing function of the average velocity gradient $\left(\nabla v_f\right)_{\mathrm{av}}$ in the flowing system. It is postulated that $\lambda$ is the local equivalent of the global parameter $\Lambda$. Therefore, the relevant characteristic time is no longer the macroscopic time $\tau$, but the microscopic time $\tau_{\mathrm{mol}} = l_{pm}/c_0$ where $l_{pm} = \left(\dfrac{M}{\mathfrak{N}_a\,\rho}\right)^{1/3}$ represents the average distance between two molecules of the fluid such that $\rho$ is the mass per unit volume, $M$ the molar mass and $\mathfrak{N}_a$ the Avogadro number. Therefore, $\lambda$ is defined such that:

$$\lambda = \left(\nabla v_f\right)_{\mathrm{av}} \tau_{\mathrm{mol}} = \frac{\left(\nabla v_f\right)_{\mathrm{av}} l_{pm}}{c_0} \qquad (17)$$

Theoretical justification of Eq. (17) requires the introduction of new concepts that are beyond the scope of this article. This justification will be given in a forthcoming paper.

Now, this definition of $\lambda$ shows that when $\left(\nabla v_f\right)_{\mathrm{av}}$ tends to infinity, $\lambda$ tends to infinity and therefore $H_N(\nu, \lambda)$ and consequently $<\left|u^2\right|>$ tend to zero. The consequence of this result is the following: with zero or very weak velocity gradient, the size of the thermal cloud pictured by $<\left|u^2\right|>$ may, in many instances depending on the thermodynamic parameters $T$ and $\rho$, be much greater than the average distance $l_{pm}$ between particles. This means that, in fact, one thermal cloud contains many particles. Since the particles cannot be localized in their thermal cloud, this means that a significant number of particles are indistinguishable in a large volume whose size may, in some cases, be as large as the macroscopic distance $d$. This volume is related to the function $H_N(\nu, \lambda = 0)$, thus to $N_{\mathrm{eff}} \approx N \gg 1$. In this sense, one can say that the system is in a coherent state. On the contrary, when the velocity gradient is high enough, $\sqrt{<\left|u^2\right|>}$ becomes smaller than $l_{pm}$, thus the coherent volume is very small, it contains only one particle, and we have $N_{\mathrm{eff}} \sim 1$. Therefore, by increasing the velocity gradient, the fluid system goes from a "coherent" state where all fluid particles are indistinguishable in a large volume, to an "incoherent" state where all fluid particles are independent from each other. This result suggests to identify this dynamical transition to some kind of Bose-Einstein condensation, where it is the velocity gradient, thus the action temperature $T_{\mathrm{A}}$ (see Ref. 7) which is the relevant parameter, and not the thermodynamic temperature $T$ as modeled in the framework of standard statistical mechanics. This transition will be illustrated in the case of water vapor in section 3.2. This feature will also be discussed in more detail in a forthcoming paper. Another way to understand this result regarding the reduction of thermal fluctuations $<\left|u^2\right|>$ caused by velocity gradients is presented in appendix A in the form of an expression that is reminiscent of the uncertainty principle of Quantum Mechanics.

At this point, it is important to stress that $\lambda$ is an external parameter determined by the experimental conditions, mainly the imposed average velocity gradient and the intrinsic parameters of the medium via the celerity $c_0$ and the intermolecular distance $l_{pm}$. Let's remember that $c_0$ characterizes the velocity at which the shear information propagates in the medium. Strictly speaking, in a real experiment the relation $c_0 = \sqrt{K/\rho}$ is only valid in an ideal case where the no-slip condition is rigorously satisfied on all surfaces in contact with the liquid. In a less perfect case (e.g. presence of bubbles, more or less hydrophobic surface), the information will take longer to cross the system, so it is as if $c_0$ had to be replaced by an effective celerity $c_{eff} = \sqrt{K_{eff}/\rho}$ where $K_{eff}$ represents an effective shear elastic constant such that $K_{eff} = C_K K$ with $C_K < 1$. If the non-perfection surface can be pictured as region where the fluid is freely slipping, the coefficient $C_K$ is a parameter that can be interpreted as the ratio of the detached surface to the total surface.

This important practical point being specified, it appears that Eq. (17) is based on the ability to determine an average velocity gradient throughout the sample. However, in a usual viscometer experiment, the velocity gradient is generally not uniform. In a tube viscometer, it is generally strong near the walls, especially as the flow rate in the tube is high, whereas it is much lower in the rest of the sample. If the gradient depends on the point $\vec{r}$, then the same must be true for the parameter $\lambda$. But since it is the function $H_N$ that quantifies the disorder which governs all properties, and since the disorders are additive, it is possible to define a new mean value $\langle \lambda \rangle$ of $\lambda$ by an average using the function $H_N$ such that: $H_N\left(v,\langle \lambda \rangle\right) = \frac{1}{V} \int_V H_N\left(v, \lambda(\vec{r})\right) dV$. Since $H_N\left(v, \lambda\right)$ is large for small $\lambda$, the value of $\langle \lambda \rangle$ is determined by low gradient values rather than high ones. Therefore, it is the quietest regions of the sample that control the reduction of fluctuations in $<\left|u^2\right|>$, the larger the corresponding volume, the greater the contribution of these quietest regions.

The combination of Eq. (15) and Eq. (17) shows that the parameter $\lambda$ is related to the Reynolds number Re in the following way:

$$\lambda \cong \mathrm{Re}\, \frac{l_{pm}}{L} \frac{1}{H_N(v)} \qquad (18)$$

In the case of the usual viscometry experiments, the term that multiplies the Reynolds number in the right member is very small (i.e. of the order of $10^{-11}$) so the $\lambda$ values will generally be lower than $10^{-8}$. Fig. 3 shows that for these numerical values of $\lambda$, the approximation $H_N\left(v, \lambda\right) \cong H_N\left(v\right)$ is perfectly justified. In other words, in the usual viscosity measurement experiments, the generalized theory can be reduced to the elastic mode theory presented in Ref. 6.

Finally, $\lambda$ is a function of $T$ and $\rho$ through the state equations of $K(T, \rho)$ and $T_t(T, \rho)$. These state equations having been determined for water in Ref. 6, and for potassium and thallium in Ref. 8. We will explore in the following section the consequences of the present modeling for various experiments done to determine the dynamic viscosity of water. We recall that the calculation program corresponding to Ref. 6 can be freely downloaded from Ref. 9.

## 3 Application to different experiments of water viscosity measurements

In Ref. 6, it was argued without a supporting calculation that external energy could be neglected in the whole analysis, i.e. that $\lambda$ is low enough such that it can always be assumed that $H_N(v, \lambda) \cong H_N(v)$ in the elastic mode theory. It is therefore important to verify here this assertion by calculation.

The pressure-density-temperature relationships used for water are taken from Ref. 10.

### 3.1. Exploration of the liquid-like phase

In the liquid and supercritical phases of water, it was shown in Ref. 6 that most of the viscosity value is determined by the "liquid" part $\eta_l$ where the function $H_N(v)$ appears in the denominator of the expression. In these states, it is therefore important to study the changes introduced by the parameter $\lambda$. In accordance with what was shown previously with Fig. 3, it is expected that there is a $\left(\nabla v_f\right)_{av}$ value that marks the change in the behavior of $H_N(v, \lambda)$ with respect to $H_N(v)$. Fig. 4 actually shows for the particular isotherm $T = 295$ K that the "separation curve" between Eq. (8) and Eq. (10) is around a value $\left(\nabla v_f\right)_{av} \approx 1000\, \text{s}^{-1}$. For $\left(\nabla v_f\right)_{av}$ values less than a few 100 s$^{-1}$, the difference between $H_N(v, \lambda)$ and $H_N(v)$ is decades smaller than the smallest error bars associated with water viscosity measurements (i.e. there is no difference for calculations to consider $H_N(v)$ instead of $H_N(v, \lambda)$).

It can be shown that the $\left(\nabla v_f\right)_{av}$ value of this "separation curve" is almost independent of the temperature value of the isotherm and therefore the chosen isotherm for Fig. 4 can be considered as representative.

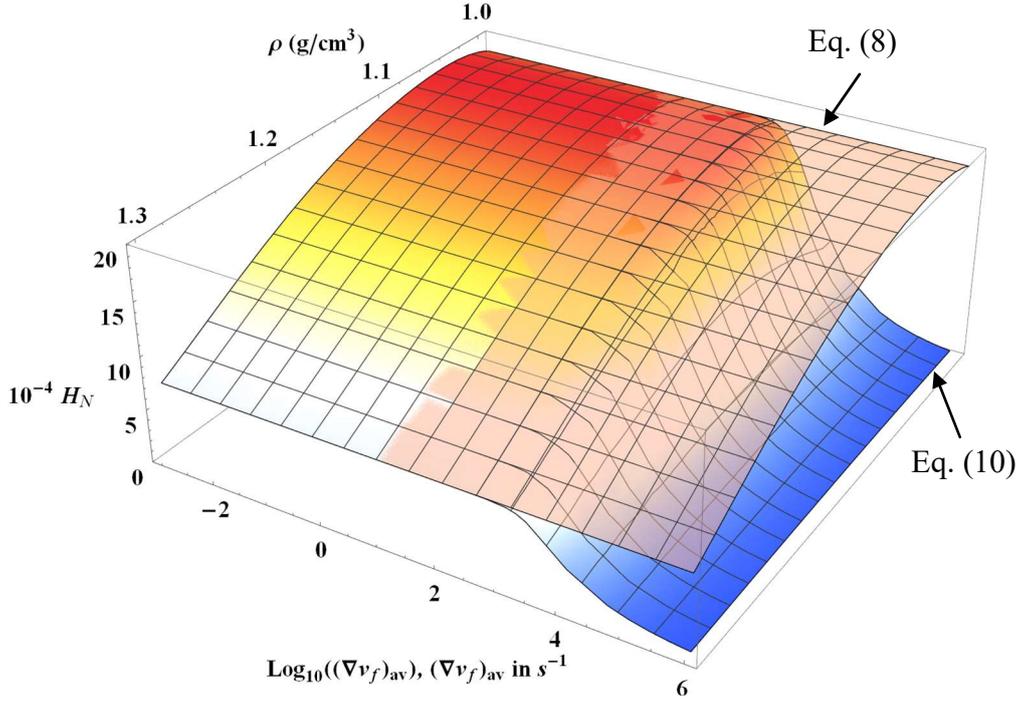

Fig. 4. Semi-logarithmic plot of the variations of Eq. (8) and Eq. (10) along the isotherm 295 K when $\overline{\lambda}$ is equal to zero or is given by Eq. (13) and Eq. (17) with water equations of state. The variation of the density scale corresponds to liquid water from its density on the saturation vapor pressure curve $\rho_{\sigma,\text{Liq}}(295\text{K}) \cong 0.99775$ g/cm$^3$ to a high density of 1.3 g/cm$^3$. The dissipative distance is $d = 100$ μm.

To conclude on the suitability of replacing $H_N(v,\lambda)$ by $H_N(v)$ for analyzing viscosity measurements, it is necessary to study some of the most constraining experiments.

The experiment of Korosi *et al.* (Ref. 11) is considered to be one of the most accurate measurements of water viscosity at atmospheric pressure between 20°C and 60°C, i.e. the corresponding uncertainty is ±0.05%. The device used for this measurement is an open-type Cannon master viscometer. In this experiment, the radius of the capillary tube is $R = 0.0165$ cm for a length $L = 45.6$ cm, but the volume of liquid in this tube is less than 1% of the total volume of liquid. Using the Poiseuille formula, the average velocity gradient on the wall of the capillary tube is: $\left|\left(\nabla v_f\right)_{\substack{\text{tube}\\\text{wall}}}\right| = \dfrac{R\,\Delta P}{2\eta\,L}$, where $\Delta P$ is the pressure difference between the top and bottom of the tube. As the viscometer works under the action of gravity, it follows that $\Delta P = \rho g L$ hence $\left|\left(\nabla v_f\right)_{\substack{\text{tube}\\\text{wall}}}\right| = \dfrac{\rho\,g\,R}{2\eta}$. In the temperature range studied by Korosi *et al.*, it is found that $\left|\left(\nabla v_f\right)_{\substack{\text{tube}\\\text{wall}}}\right|$ varies between 806 s$^{-1}$ and 1706 s$^{-1}$. According to Fig. 4, it can be seen that these values are distributed around the "separation curve" of Eq. (8) and Eq. (10). But these values represent a maximum value of the gradient in the sample. Now, it was mentioned earlier that $\left(\nabla v_f\right)_{\text{av}}$ is essentially determined by the quietest flow regions, i.e. here by what happens in the water tank which has a radius about 100 times larger than the capillary tube. It is deduced that the required average $\left(\nabla v_f\right)_{\text{av}}$ is such that $\left(\nabla v_f\right)_{\text{av}} \approx \left|\left(\nabla v_f\right)_{\substack{\text{tube}\\\text{wall}}}\right|\Big/(100)^3$, i.e. the value to be considered for this experiment is of the order $10^{-3}$ s$^{-1}$ (or a value of $\lambda \approx 1.8 \times 10^{-16}$).

Fig. 4 shows that for such a value, there is no calculable difference between $H_N(v)$ and $H_N(v,\lambda)$.

As a second example, we will consider Hallett's experiment (Ref. 12) which determines the viscosity of water always at atmospheric pressure but in the supercooled phase between -24°C and 0°C. Hallett's experiment corresponds, as before, to observing the flow in a capillary tube with a radius of $R = 0.01$ cm and length $L = 15$ cm. In this experiment the pressure difference in the capillary tube is set at $\Delta P = 1167.13$ Pa. In this temperature range, $\left|\left(\nabla v_f\right)_{\substack{\text{tube}\\\text{wall}}}\right|$ varies between 70 s$^{-1}$ and 217 s$^{-1}$. But as in the experiment of Korosi $et\ al.$, one must consider what happens in the water tank, which has a radius 200 times larger than that of the capillary tube. It is deduced that the average $\left(\nabla v_f\right)_{\text{av}}$ is such that $\left(\nabla v_f\right)_{\text{av}} \approx \left|\left(\nabla v_f\right)_{\substack{\text{tube}\\\text{wall}}}\right|/(200)^3$, of the order $10^{-5}$ s$^{-1}$ (or a value of $\lambda \approx 1.8{\times}10^{-18}$). Here again, it can be considered that there is no difference between $H_N(v)$ and $H_N(v,\lambda)$.

We will analyze a third example by considering the experiment of Först $et\ al$. (Ref. 13) called "optical method" which consists in observing the fall of a spherical glass ball inside a tube. The ball has a radius $R_S = 25.4$ μm and a density $\rho_S = 2.42$ g/cm$^3$. The analysis of these data in Ref. 6 showed that the dissipative distance $d = R_S$ should be considered here. It is assumed that a maximum value of the velocity gradient on the ball surface can be determined from the following formula: $\left|\left(\nabla v_f\right)_{\substack{\text{ball}\\\text{surface}}}\right| = \dfrac{2R_S(\rho_S - \rho)g}{9\eta}$. In the studied pressure range, $\left|\left(\nabla v_f\right)_{\substack{\text{ball}\\\text{surface}}}\right|$ varies between 51 s$^{-1}$ and 77 s$^{-1}$ along the isotherm at 293.15 K. If we simply assimilate these values to $\left(\nabla v_f\right)_{\text{av}}$, it is found that the relative deviation between $H_N(v)$ and $H_N(v,\lambda)$ varies from $5{\times}10^{-6}$ % to $3.5{\times}10^{-7}$ %, which is totally negligible on the calculation of viscosity given that the uncertainty of the data is here $\pm5$%. Now the diameter of the tube is 27.6 times larger than the diameter of the ball, therefore if we take it as the diameter of an effective water tank then we obtain $\left(\nabla v_f\right)_{\text{av}} \approx 3{\times}10^{-3}$ s$^{-1}$ (or a value of $\lambda \approx 5{\times}10^{-16}$). This leads to the same conclusion as in the previous examples.

We will consider a fourth and last example with the analysis of the experiment of Mariens $et\ al$. (Ref. 14) using the oscillating disc method. By considering the solution of the Navier-Stokes equation for region II as defined in Ref. 14, one can directly deduce an order of magnitude of the mean velocity gradient such that: $\left(\nabla v_f\right)_{\text{av}} \approx \dfrac{2\pi\,\Delta}{T}\,\alpha(0)\sqrt{\dfrac{\rho}{\eta}\dfrac{2\pi\,\Delta}{T}}\,R_d$, where here $T$ is the period of the oscillation in the fluid, $2\pi\Delta$ the logarithmic decrement of the amplitude in the fluid, $\alpha(0)$ the angular deflection of the disc at instant $t = 0$ and $R_d$ the disc radius. From the data of their Table III (Ref. 14) for water at atmospheric pressure, it can be deduced for the studied temperature range that $\left(\nabla v_f\right)_{\text{av}}$ varies between $4.3{\times}10^{-3}$ s$^{-1}$ and $2.1{\times}10^{-2}$ s$^{-1}$. It is then deduced that the relative deviation between $H_N(v)$ and $H_N(v,\lambda)$ varies from $5{\times}10^{-14}$ % to $4{\times}10^{-12}$ %, which is totally negligible on the calculation of viscosity.

Thus, through these examples it appears that for the analysis of the usual viscosity experiments of liquid water, one can well consider that the perturbation by the measurement is negligible, thus replacing Eq. (10) by Eq. (8) in the elastic mode theory is an allowed simplification.

### 3.2. The dilute-gas limit

The dilute-gas limit is generally associated with the perfect gas behavior, so it is important to study the behavior of the present modeling within this limit. Expressions for the various model parameters in the dilute-gas limit are given in Ref. 6, sections III.B and IV.B.4. Here, we are interested in the expression for the size of the thermal cloud because, in this limit, $\sqrt{<|u^2|>}$ can become comparable to the dissipative distance $d$ in a real experiment.

The expression for $<|u^2|>$ given by Eq. (6) is written as the product of two terms. The first term is $k_B T\, q_c /(\pi^2 K)$. In the dilute-gas limit, $\lim_{\rho \to 0} K \propto \rho^3$ and $\lim_{\rho \to 0} q_c \propto \rho^{-1}$ so that the first term varies as $\rho^{-4}$. For the second term, **(i)** if $\bar{\lambda}$ is sufficiently small, i.e. satisfies condition (11), then one has $\lim_{\rho \to 0} v \to 2$ and $\lim_{\rho \to 0} H_N(v,\lambda) \cong H_N(2,0) = N-1 \propto \rho^2$. In total, in the weak perturbation limit, $<|u^2|>$ diverges as $1/\rho^2$. In practice, this divergence of $<|u^2|>$ is limited by the finite sample size, comparable to the fluctuative distance $d_N$; **(ii)** If instead $\bar{\lambda}$ is sufficiently large, according to Eq. (9), $H_N(v,\lambda) \propto (1/\lambda^3)(1-1/N^3)$. Since according to Eq. (17), $\lim_{\rho \to 0} \lambda \propto \rho^{-\frac{4}{3}}$, and since $\lim_{\rho \to 0} (N-1) \propto \rho^2$, then in first approximation $\lim_{\rho \to 0} H_N(v,\lambda) \propto \rho^6$. In total, for $\rho \to 0$, in the strong perturbation limit, $<|u^2|>$ is proportional to $\rho^2$.

As mentioned in section 2.2, this analysis confirms the existence of a Bose-Einstein like transition in the case of gas. Fig. 5 illustrates this behavior in the case of steam at room temperature. The transition is clearly visible for sufficiently low densities. The transition is smooth and occurs around an average velocity gradient of about 100 s$^{-1}$. It is then less and less visible as the density increases.

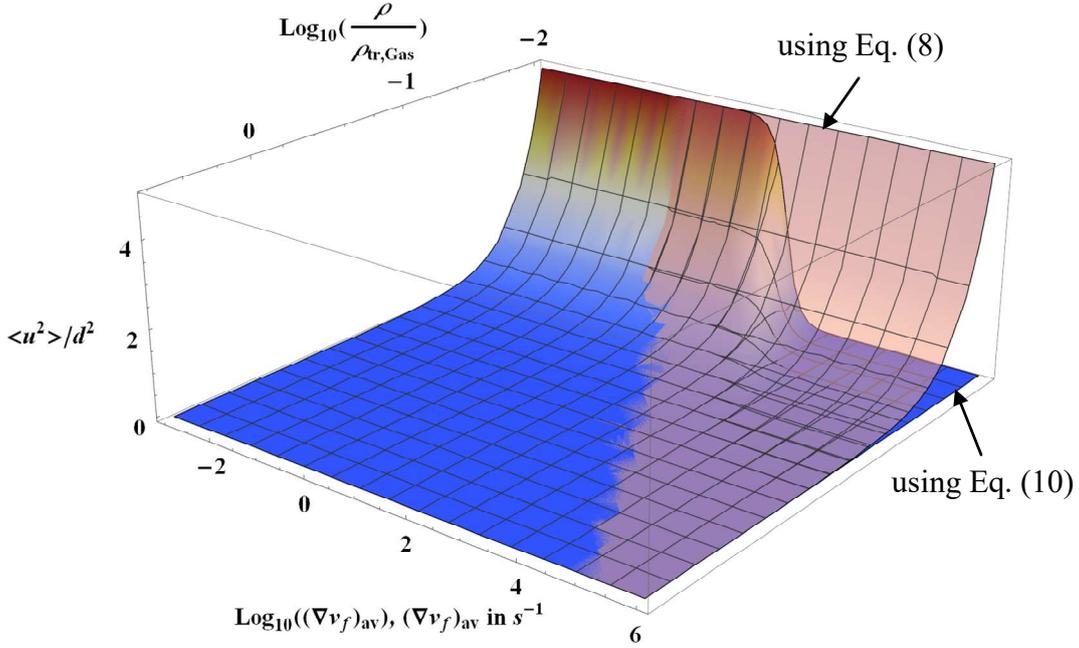

Fig. 5. Semi-logarithmic plot of the variations of Eq. (6) along the isotherm 295 K when $\lambda$ is equal to zero or is given by Eq. (13) and Eq. (17) with water equations of state. The variation of the density scale corresponds to the steam from its density on the saturation vapor pressure curve $\rho_{\sigma,\text{Gas}}(295\text{K}) \cong 1.927 \times 10^{-5}$ g/cm$^3$ to $\rho_{\text{tr,Gas}}/100$, where $\rho_{\text{tr,Gas}}$ represents the triple point gas density of water. The dissipative distance is $d = 100$ μm.

Fig. 5 also shows that as long as $(\nabla v_f)_{\text{av}}$ is small enough then calculating $<|u^2|>$ with Eq. (8) or Eq. (10) is strictly equivalent. The difference between the two expressions occurs for extremely small densities: for $d = 100$ μm, the condition $<|u^2|>/d^2 = 1$ is reached for a density much smaller than $\rho_{\text{tr,Gas}}$ and is practically independent of $(\nabla v_f)_{\text{av}}$. It can be observed that the increase in $(\nabla v_f)_{\text{av}}$ eventually combines with density variation in such a way that $<|u^2|>$ remains very small.

In the gas phase, it has been shown in Ref. 6 that the essential part of viscosity variation is determined by the Knudsen term $\eta_{Knu}$ which is not dependent on the $H_N$ function. However, the liquid term $\eta_l$ represents about half of the total viscosity value and therefore, in order to validate this analysis, it is important to show that $H_N(v,\lambda)$ does not differ significantly from $H_N(v)$ in the corresponding experimental conditions.

We will take here as an example the experiment of Yasumoto (Ref. 15) because these measurements correspond to the lowest densities reached in the vapor phase of water. This experiment consists in measuring the vapor viscosity by means of a tandem capillary-flow viscometer whose capillary radius is about $R = 0.06$ cm. A short capillary serves as a "measuring capillary" while a second capillary 2.43 times longer connected in series allows the laminar flow rate to be reached for the measurement. Knowing the flow rate $Q$ for each measuring point, one can determine the maximum velocity gradient on the wall of the capillary tube, such as $\left|(\nabla v_f)_{\text{wall}}^{\text{tube}}\right| = \dfrac{16\,Q}{\pi\,R^3}$. For Yasumoto's temperature and density ranges, a value between $4.5 \times 10^4$ s$^{-1}$ and $1.476 \times 10^5$ s$^{-1}$ is obtained. Fig. 6 shows that if we assign these values to $\left(\nabla v_f\right)_{\mathrm{av}}$ then we are in the region where Eq. (10) varies strongly with the velocity gradient parameter (on the other hand, it can be observed that the function $H_N$ does not vary with temperature). As was done in the previous section, a $\left(\nabla v_f\right)_{\mathrm{av}}$ value is determined here using the ratio of the lengths of the two capillary tubes. It is then deduced that $\left(\nabla v_f\right)_{\mathrm{av}}$ varies between 3136 s$^{-1}$ and 10 286 s$^{-1}$. The relative deviation between $H_N(v)$ and $H_N(v, \lambda)$ varies from $2 \times 10^{-3}$ % to $7 \times 10^{-2}$ %, which is totally negligible on the calculation of viscosity given that the uncertainty of the viscosity data is here $\pm 1\%$.

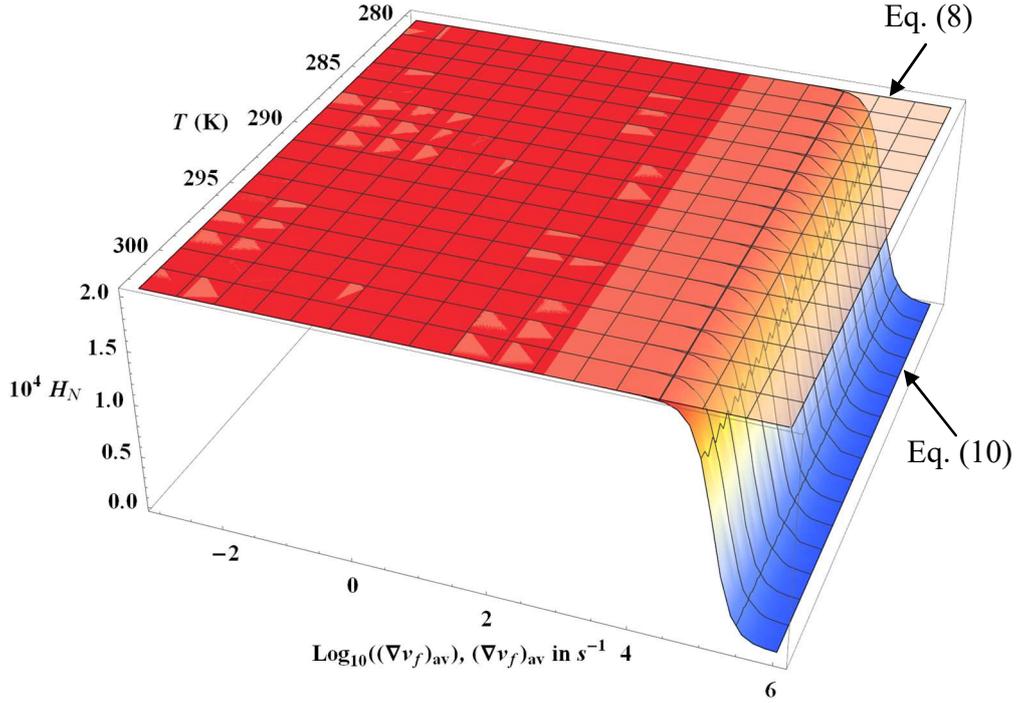

Fig. 6. Semi-logarithmic plot of the variations of Eq. (8) and Eq. (10) along the isochor $6.1036 \times 10^{-6}$ g/cm$^3$ when $\lambda$ is equal to zero or is given by Eq. (13) and Eq. (17) with water equations of state. The temperature range corresponds to that explored by Yasumoto's experiment (Ref. 15). The dissipative distance is $d = 100$ μm.

To complete the previous analysis, let us consider one of the "worst" case where $\left(\nabla v_f\right)_{\mathrm{av}} = 10^4$ s$^{-1}$. Fig. 7 shows the deviation that this $\left(\nabla v_f\right)_{\mathrm{av}}$ value produces on the $H_N$ function. It is observed that the deviation only starts to become significant in relation to the experimental uncertainties in the region near the critical point. It can be concluded that it is possible to use Eq. (8) instead of Eq. (10) for all usual viscometry experiments with a vapor density lower than 0.05 g/cm$^3$.

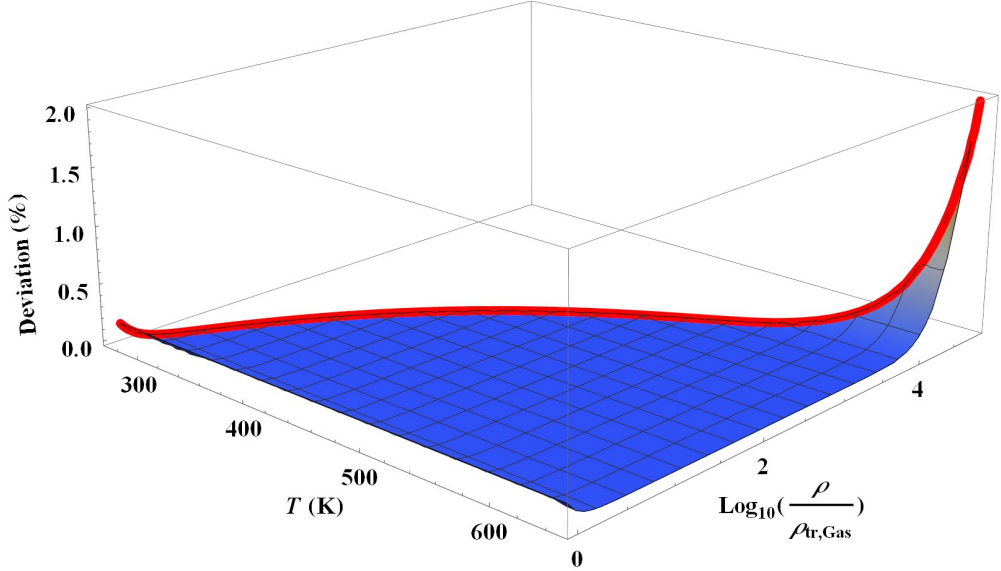

Fig. 7. Deviation plot $100\left(H_N(v) - H_N(v,\lambda)\right)/H_N(v)$ for steam. The water vapor density range is from $\rho_{\text{tr,Gas}}$ to the critical density $\rho_c$ and the temperature range is from the triple point temperature $T_t$ to the critical temperature $T_c$. The red thick curve represents the vapor coexistence curve of water. The dissipative distance is $d = 100$ µm.

From Eq. (17), the parameter $\lambda$ can be also interpreted as the ratio of two velocities such that $\lambda = \bar{v}_m/c_0$ where $\bar{v}_m$ represents an average relative velocity between two neighboring atoms (or molecules) of the medium. In all cases, $\bar{v}_m$ cannot exceed the speed of light $c$ (i.e. the velocity gradient cannot exceed $c/l_{pm}$), therefore $\lambda$ admits a limit value $\lambda_{\max} = c/c_0$. However in the dilute-gas limit, we have shown in Ref. 6 that $\lim_{\rho \to 0} c_0 \propto \rho$. It follows that $\lambda_{\max}$ (and therefore $\lambda$) can become infinite in the limit $\rho \to 0$. In other words, $N_{\text{eff}}$ tends to 1 in this limit. This still implies that the particles can only be considered as isolated in the limit $\rho \to 0$. However, it should be noted that the divergence of $\lambda$ is theoretical since from an experimental point of view, one cannot reach $\rho = 0$. Thus, in practice the divergences do not exist but only result in large numbers. Finally, it should be noted that in the limit $\rho \to 0$, the cutoff pulsation $\omega_c$ of the inertial mode theory tends to zero as $c_0$. In other words, there exists in this limit only the transient regime which persists for an infinitely long time (i.e. the inertial modes are irrelevant).

### 3.3. Experiments from low to high shear rates at atmospheric pressure and room temperature

In most of the previously analyzed experiments the velocity gradient varies strongly at various points in the system and therefore the evaluation of the average gradient is not easy. It is then useful to analyze more particularly the viscosity data in experiments where the gradient can be easily determined and moreover is quasi-uniform in the studied system. To carry out such experiments two rotating experimental devices were used and their main characteristics are presented in the next section.

### 3.3.1. Presentation of devices and experimental results

The first set-up used is a HAAKE DC60/1 device with a double cone rotor. Some geometrical characteristics of the rotor are grouped in Table 1. The tank containing the liquid is temperature-controlled to an accuracy of 0.1 K. The rotor and the tank are made of stainless steel.

| Cone diameter (mm) | Bottom cone angle (deg) | Upper cone angle (deg) | Bottom cone truncation (mm) | Height of the cylindrical shape separating the two cones (mm) |
|---|---|---|---|---|
| 63.562 | 1.023 | 5 | 0.054 | 2 |

Table 1. Some geometrical features of the HAAKE DC60/1 rotor.

Depending on the filling of the tank, it is possible to shear the liquid only with the bottom cone (i.e. this kind of experiment will be called "simple cone") or with both cones (and this kind of experiment will be called "double cone"). Moreover, with this instrument it is possible to impose a linear stress ramp or a linear strain ramp as function of time.

Fig. 8 shows three regions where the viscosity behavior with the velocity gradient is significantly different: subregion 1 corresponds to the medium shear during the dynamic transition phase; in subregion 2, the viscosity is independent of the velocity gradient and thus corresponds to the value of the viscosity in the Newtonian regime; finally, subregion 3 shows a regular increase in viscosity with an inflection point in the vicinity of 2000 s$^{-1}$. This qualitative behavior is not specific to the double cone experiment but is also observed with the simple cone experiment as well as with the second experimental device.

Fig. 8 also shows that the results for liquid water in subregions 2 and 3 do not depend on the imposed linear ramp type; therefore, no distinction will be made between the results later on.

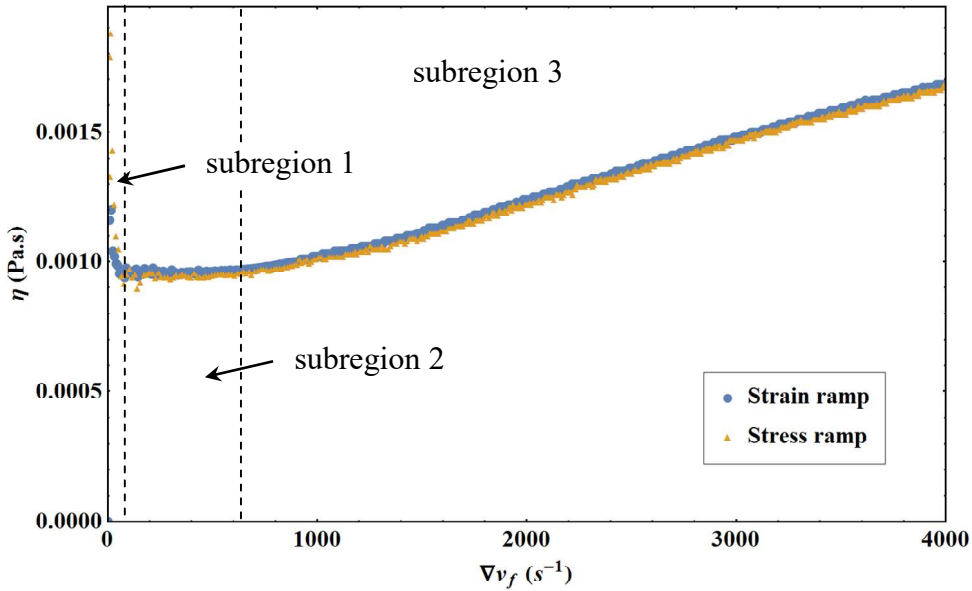

Fig. 8. Experimental results of liquid water viscosity as a function of the velocity gradient obtained with the double cone by realizing either a linear stress ramp or a linear strain ramp. $T$ = 293.15 K.

The second experimental device used is a Couette cell whose inner cylinder is fixed and the outer cylinder rotates at a constant velocity in order to prevent the occurrence of the Taylor-Couette instability. The inner fixed cylinder ends with a conical shape at its base corresponding to a thickness of 1 mm. Some geometrical characteristics of the Couette cell are grouped in Table 2. The cell temperature is not regulated here and is at room temperature (i.e. ~293.15 K). It should be noted that the cell is made of polycarbonate which is a weakly hydrophilic material. With this cell, we can only impose a linear rotation velocity ramp of the external cylinder as function of time, which is equivalent to imposing a linear strain ramp as function of time.

| Inner cylinder diameter (mm) | Outer cylinder diameter (mm) | Height of the cell (mm) | Height of water (mm) |
|---|---|---|---|
| 20 | 21 | 40 | 16.787 |

Table 2. Some geometrical features of the Couette cell.

The volumes of water used in the double cone experiment and in the Couette cell are comparable. In the simple cone experiment, the volume is slightly lower than in the other two experiments. The characteristic values of these volumes are grouped in Table 3. It should also be noted that the height of water in the Couette cell is much lower than the cylinders height in order to prevent liquid ejection during rotation.

| Double cone water volume (cm$^3$) | Simple cone water volume (cm$^3$) | Couette cell water volume (cm$^3$) |
|---|---|---|
| ~ 3.2 | ~ 2 | 3 |

Table 3. Characteristic water volumes used in the different experiments.

Fig. 9 shows that whether it is for the simple cone or the Couette cell, the three subregions described in Fig. 8 are observed. However, it can be noticed that the numerical values in subregion 3 are significantly higher with the simple cone and with the Couette cell than with the double cone. Moreover, "stalls" appear in subregion 3 at high shear rates. These differences are partly related to the presence of a free surface in the simple cone and Couette experiments contrary to the case of the double cone experiment. But the high numerical values obtained with the Couette cell are also due to the cell material which causes a strong slip of the liquid as we shall see.

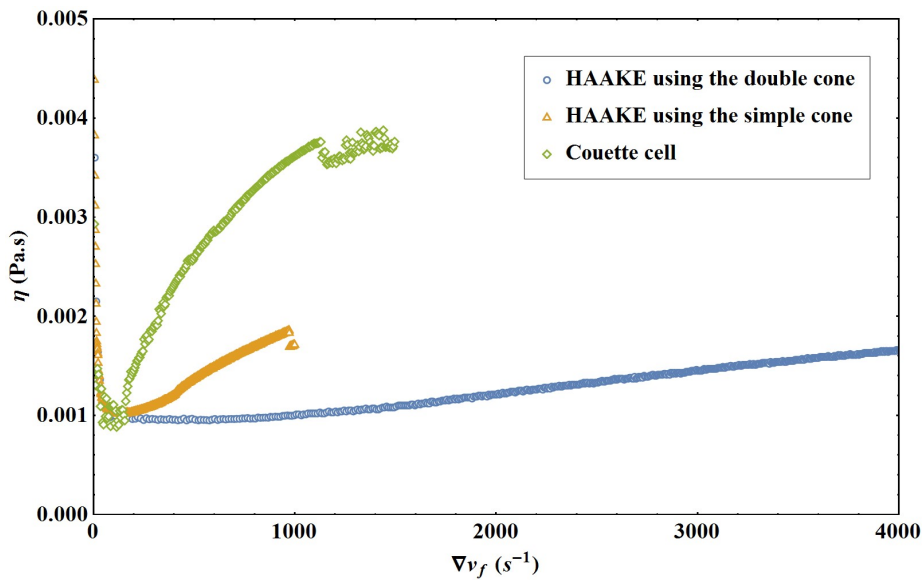

Fig. 9. Experimental results of liquid water viscosity as a function of the velocity gradient obtained with the different experimental devices used. $T$ = 293.15 K.

Considering that the experimental results analyzed in the different devices involve linear strain ramps, it is useful to give some explanation on the measurements. A linear strain ramp is equivalent to imposing a linear ramp on the rotor angular velocity in the different experiments. The theoretical models established then allow to transform this angular rotation velocity into a velocity gradient (e.g. see Ref. 1). The torque exerted on the rotor is then measured and converted into stress from these theoretical models. The viscosity is then simply defined as the ratio of the stress to the velocity gradient. Table 4 regroups the characteristic values of the linear strain ramp slopes imposed for the different types of data presented in Fig. 9. These values make it easy to convert the values of the velocity gradients into time and vice versa. The second line of Table 4 allows to check the agreement between the slopes of the strain ramps and the characteristic durations of the experiments.

|  | Double cone | Simple cone | Couette cell |
|---|---|---|---|
| Slope of the linear strain ramp (s$^{-2}$) | 1.6148 | 0.9989 | 1.6715 |
| Experiment duration (s) | 2500 | 1000 | 900 |

Table 4. Characteristic values of the linear strain ramp slopes for the different experimental data shown in Fig. 9.

Ultra-pure deionized water was used for the experiments presented in the paper. However, tests with different qualities of deionized water up to the use of tap water did not show any change in the observation of the different regimes as well as their variations. This is perfectly in line with the viscosity data for salt water that have been analyzed in Fig. 56 of Ref. 6 and that show that the viscosity variations can be described in the same way as for liquid water. In other words, the quality of the water does not matter for the variations, and it is only the set of absolute values that are numerically shifted.

In all cases, the experimental conditions correspond to water at atmospheric pressure and a mean temperature of 293.15 K. For these conditions, the 1995 IAPWS state equation formulation (Ref. 10) gives a liquid water density $\rho = 0.998207$ g/cm$^3$ and the 2008 IAPWS formulation (Ref. 16) gives a liquid water viscosity $\eta = 1.0016$ mPa.s. The elastic mode theory (Ref. 6) gives a liquid water viscosity $\eta = 1.0018$ mPa.s with a Knudsen term $\eta_{Knu} = 0.0379$ mPa.s. However, the apparatus constants in these experimental devices were set to find $\eta = 1.002$ mPa.s. This said, the absolute values given by the experimental devices vary quite strongly while the variations of viscosity as a function of the velocity gradient can be always superimposed for a given type of experiment. Changing the rise time by a factor 2 to 4 leads to the same results to experimental precision.

Knowing the experimental conditions, the characteristic times of different processes can be determined. Table 5 shows that the shortest characteristic time is the time $\tau$ related to the propagation of the shear information. The viscous diffusion time $\tau_v$ is several orders of magnitude higher than the stress relaxation time. This stress relaxation time for water is deduced from Ref. 7 for deformations higher than 60% which is always the case in the experiments performed here. Table 5 shows that the system has time to reach the steady state for each measurement made.

| $\tau$ (s) | $\tau_v \approx \tau\, H_N(v)$ (s) | Stress relaxation time (s) | Interval between two experimental points (s) |
|---|---|---|---|
| $10^{-8}$ | $10^{-2}$ | $10^{-1}$ | $10^{0}$ |

Table 5. Characteristic times values for the two rotating experimental devices with liquid water.

### 3.3.2. Experimental results analysis

The usual rheological models introduce non-dimensional numbers such as the Weissenberg number or the Reynolds number. But in this new approach, other non-dimensional numbers are relevant such as the parameter $\lambda$ which represents the ratio between the injected mechanical energy and the energy of the elastic response of the system or $T_A^*$ which represents the ratio between the action temperature and the re-action temperature (see Ref. 7). If some of these dimensionless numbers can be related to usual dimensionless numbers, they should not be confused with them.

We will start by analyzing subregion 3 because it does not require to calculate the viscosity absolute value in the Newtonian regime. One can simply take the experimental value corresponding to subregion 2, whatever its value. Indeed, if we note $\eta_0$ the value of the viscosity in subregion 2 corresponding to the usual Newtonian regime for which it has been shown previously that the viscosity can be described using $H_N(v)$ only, then the influence of $\lambda$ on the viscosity value is to increase it with respect to the Newtonian regime in such a way that:

$$\eta(\lambda) = \eta_0 \left( 1 - \frac{\eta_{Knu}}{\eta_0} \right) \frac{H_N(v)}{H_N(v, \lambda)} + \eta_{Knu} \qquad (19)$$

where $H_N(v)$ is given by Eq. (8) and $H_N(v, \lambda)$ by Eq. (10). The viscous term $\eta_{Knu}$ represents the gas-like term (also called the Knudsen term) corresponding to the contribution of the gas released by the action of the shear stresses. The parameters $v$ and $\eta_{Knu}$ are here two constants which depends only on temperature and density of the liquid. The value of $\eta_0$ is determined experimentally from subregion 2. The quantity $\lambda$ being the variable, Eq. (19) is then entirely determined if $N$ is known. Remember that $N$ is decomposed as follows in Ref. 6:

$$N - 1 = d_N \frac{q_{c0,\text{crit}}}{2\pi} \qquad (20)$$

where $q_{c0,\text{crit}} = q_{c0}(\rho_c) = \left( \frac{6\pi^2 \rho_c \, \mathfrak{R}_a}{M \, n_B} \right)^{1/3}$ represents the cut-off wave vector modulus of the elastic modes at the critical density $\rho_c$ and $d_N$ represents the fluctuative distance whose value depends on some characteristic length of the experimental set-up. The distance $d_N$ is therefore the only adjustable parameter of the model.

Fig. 10a shows that Eq. (19) makes it possible to reproduce very faithfully the evolution of viscosity in the whole subregion 3 by fixing $d_N$ at the value of 1.3144 cm. The value of $d_N$ is slightly lower than the cubic root of liquid water volume in this experimental run, i.e. 1.4342 cm. This value of $d_N$ is perfectly in line with what is expected for this type of experimental device.

Fig. 10b shows in a different form the comparison between the experimental data and Eq. (19). It can be seen that the deviation is within an uncertainty band of $\pm 1.5\%$. This uncertainty corresponds to that of the experimental device which is about $\pm 2\%$. Then, it can be noticed an oscillation of great "wavelength" of the uncertainty. This oscillation is in phase with the oscillation of the temperature control ($\pm 0.05$ K) and is therefore inherent to the device itself.

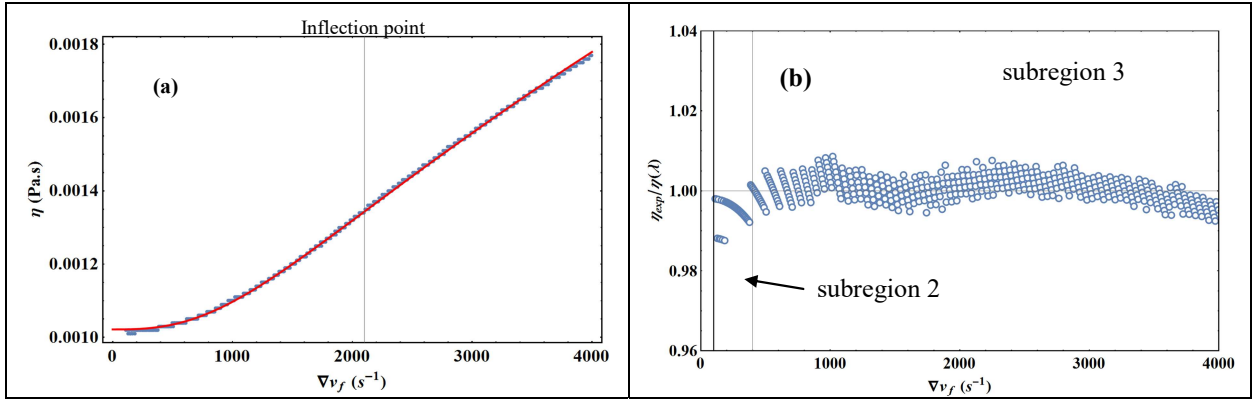

Fig. 10. (a) Comparison of theoretical model (red curve, Eq. (19) with $d_N = 1.3144$ cm ) with the experimental results (blue points) for liquid water viscosity as a function of the velocity gradient. (b) Deviation of the experimental data with Eq. (19). Only the experimental points corresponding to subregions 2 and 3 are represented. Double cone experiment at $T = 293.15$ K.

It was mentioned in the previous section that the viscosity evolution in subregion 3 admitted an inflection point. Fig. 11 shows very clearly the existence of this inflection point at 2100.13 $s^{-1}$ and indicates a positive curvature of the viscosity variation for lower velocity gradients while the curvature is negative for high values of the velocity gradient. One has the feeling that the curve tends asymptotically to zero when the velocity gradient becomes very large. From a theoretical point of view, given Eq. (9), Eq. (19) is written in the limit:

$$\eta\left(\lambda \rightarrow \infty\right) = 3\eta_0\left(1 - \frac{\eta_{Knu}}{\eta_0}\right)H_N\left(v\right)\lambda^{1+v} + \eta_{Knu} \qquad (21)$$

The second derivative of Eq. (21) is such that:

$$\frac{d^2\eta}{d\lambda^2}\left(\lambda \rightarrow \infty\right) = 3v\left(1+v\right)\eta_0\left(1 - \frac{\eta_{Knu}}{\eta_0}\right)H_N\left(v\right)\lambda^{v-1} \qquad (22)$$

Thus, when $v > 1$, the second derivative is positive at the limit and tends to infinity. One can verify that another inflection point appears for a velocity gradient value of the order of $6.82 \times 10^{11}$ $s^{-1}$, using the parameters of Fig. 10. In practice, for this velocity gradient value the turbulence has already appeared and the present modeling is no longer valid (see appendix B).

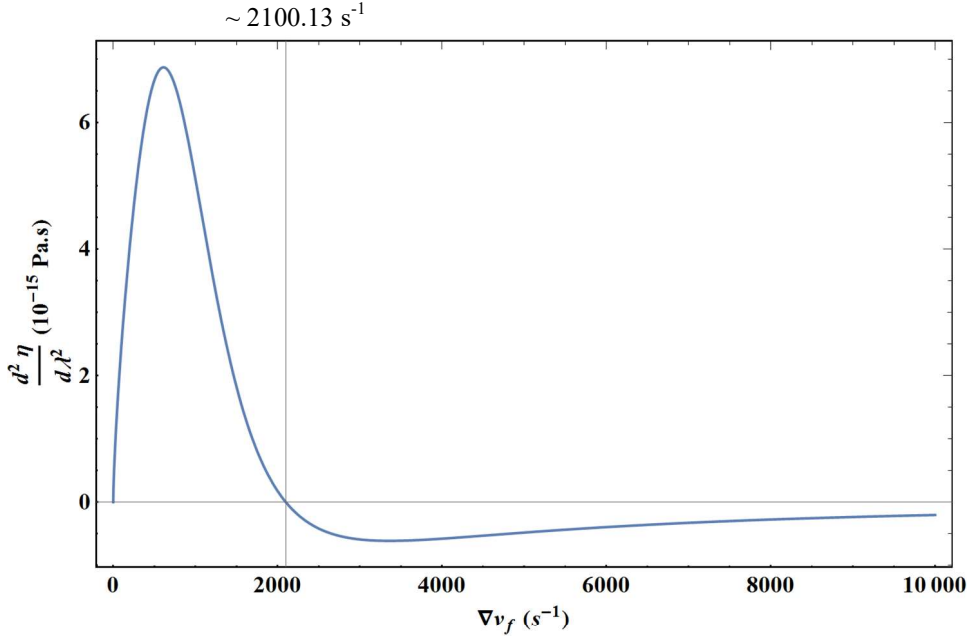

Fig. 11. Second derivative of Eq. (19) as function of the velocity gradient using the parameters corresponding to Fig. 10.

Until now, the experimental data analyzed corresponded to those for which the contact of the liquid with the surfaces could be considered as perfect (i.e. no slipping at wall). Indeed, in the double cone the wall surfaces could be considered sufficiently hydrophilic (i.e. almost perfect wetting) and there is no free liquid surface. The data concerning the experiments with the simple cone and the Couette cell on the contrary involve free liquid surfaces as well as partial slipping at the walls. Therefore, as mentioned in section 2.2, in order to analyze these data, the elastic constant $K$ must be replaced by the effective elastic constant $K_{eff} = C_K K$ in the various theoretical expressions. On the other hand, the intrinsic parameters that define the medium at equilibrium, such as the transition temperature $T_l$, are not impacted by the coefficient $C_K$. Specifically the value of exponent $v$ which appears in all calculations is not affected by this replacement.

Given the transformations of $K$ into a $K_{eff}$, the data in Fig. 9 corresponding to the simple cone experiment can be analyzed similarly to those from the double cone experiments. First of all, it can be seen in Fig. 12a that the plateau corresponding to subregion 2 is strongly shifted in absolute value. Moreover, this plateau is rather narrow. The theoretical curve determined with Eq. (19) shows that the beginning and the end of subregion 3 can be reproduced correctly, while in the middle appears as a bump. Fig. 12b shows that the evolution of the viscosity is only a consequence of the stress variation. On this same figure, it can be seen that the stress has a significant change of evolution around 400 s⁻¹ and then suddenly returns to a "normal" behavior beyond 970 s⁻¹. If the bump is ignored for the moment, the theoretical curve requires that the parameter $d_N$ be equal to the cone radius. This is perfectly consistent with what was found in Ref. 7 with the case of the plate-plate rheometer which also contained a free surface. The presence of this free surface also has the effect of inducing wall slip which is translated in the elastic mode theory by a decrease in the static shear elastic constant $K$ value. Indeed, the theoretical curve requires to decrease $K$ by a coefficient $C_K = 0.7655$. This is also in perfect agreement with what has been shown in Ref. 7.

Let us now analyze the case of the bump in subregion 3. The bump appears because the stress is stronger than it should be, that is, the torque to drive the liquid is too strong. This suggests an "abnormally" strong slip. Indeed, Fig. 12c shows that this bump can be reproduced by changing only the value of the coefficient $C_K$, which must decrease until it reaches a value 1.49 times lower. The relatively sharp decrease of the viscosity during the stalls observed at high shear rate in this simple cone experiments can be interpreted as a reduction of the slip for certain values of the rate, phenomena probably related to the existence of a free surface for the liquid combined with the unavoidable vibrations associated with relatively high velocities of rotating mechanical systems.

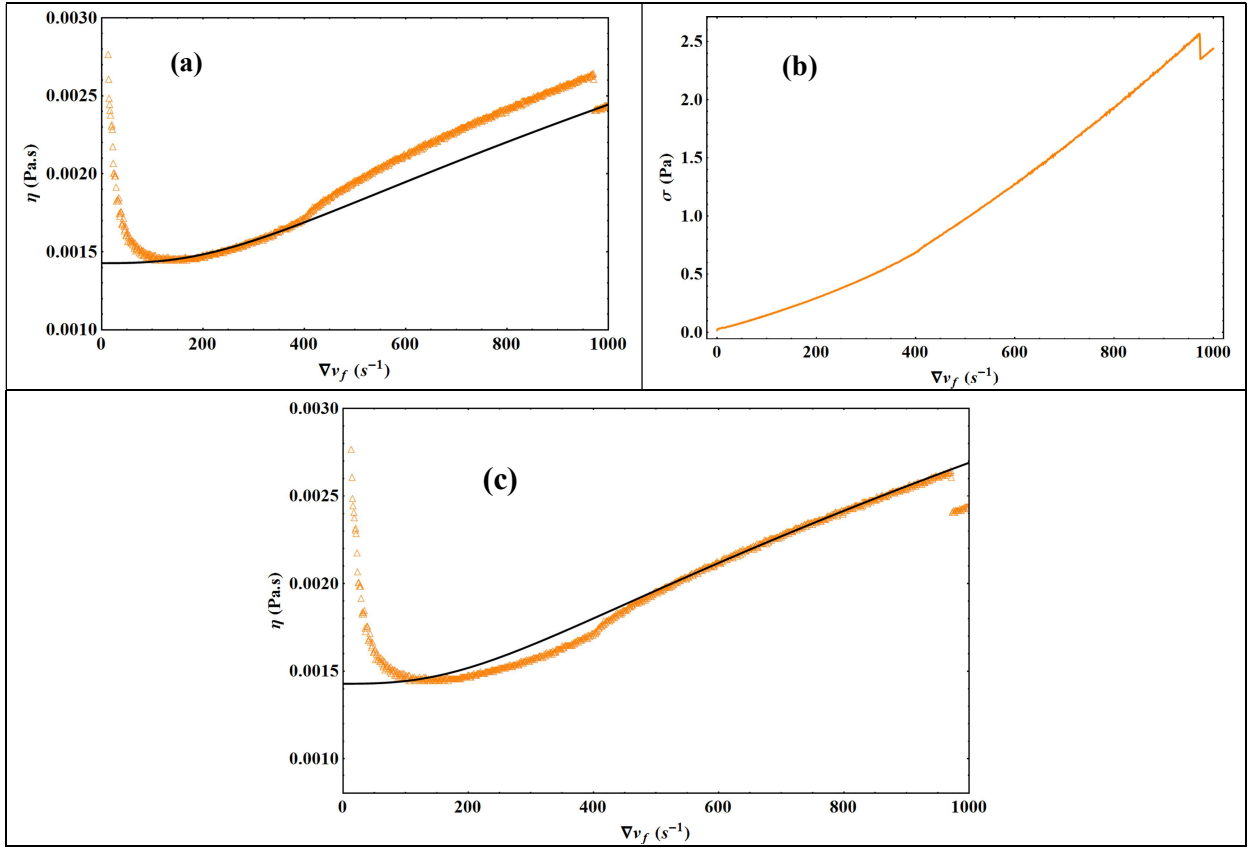

Fig. 12. Liquid water experiment at $T$ = 293.15 K with the HAAKE simple cone by imposing a linear strain ramp. (a) Experimental results (orange triangles) versus theoretical model (black curve with $d_N$ = 3.1781 cm and $C_K$ = 0.7655) for the viscosity as a function of the velocity gradient. (b) The experimental stress versus the velocity gradient. (c) Experimental results (orange triangles) versus theoretical model (black curve with $d_N$ = 3.1781 cm and $C_K$ = 0.5137) for the viscosity as a function of the velocity gradient.

The viscosity variation in Fig. 9 corresponding to the Couette cell experiment has the same qualitative behavior as the simple cone experiment. This behavior must therefore be explained in the same way. First of all, it can be observed in Fig. 13 that the data here are much less precise. It is interesting to note that the theoretical curve in Fig. 13a imposes a value of $d_N$ corresponding to the height of water in the cell. Also, a lower value of the static shear elastic constant $K$ must be introduced. This is in accordance with the fact that there is the presence of a free surface as in the simple cone experiment which induces a slip but it appears an intensification of this slip because of the more hydrophobic material of the Couette cell compared to the stainless steel of the cone. The value of $C_K$ is perfectly compatible with the results of Badmaev *et al.* (Ref. 17) to reproduce the values of the liquid water shear elastic modulus $G'$ for low wettability of the contact surface (see the corresponding discussion in Ref. 7).

As with the simple cone, Fig. 13b shows that the bump in subregion 3 is due to a significant change in stress variation between 155 s$^{-1}$ and 1130 s$^{-1}$. Again, this bump represents an "abnormally" strong slip. Fig. 13c shows that this bump can be reproduced by only decreasing the value of the coefficient $C_K$ until it reaches a value 1.76 times lower. This analysis is therefore consistent with that of the simple cone experiment.

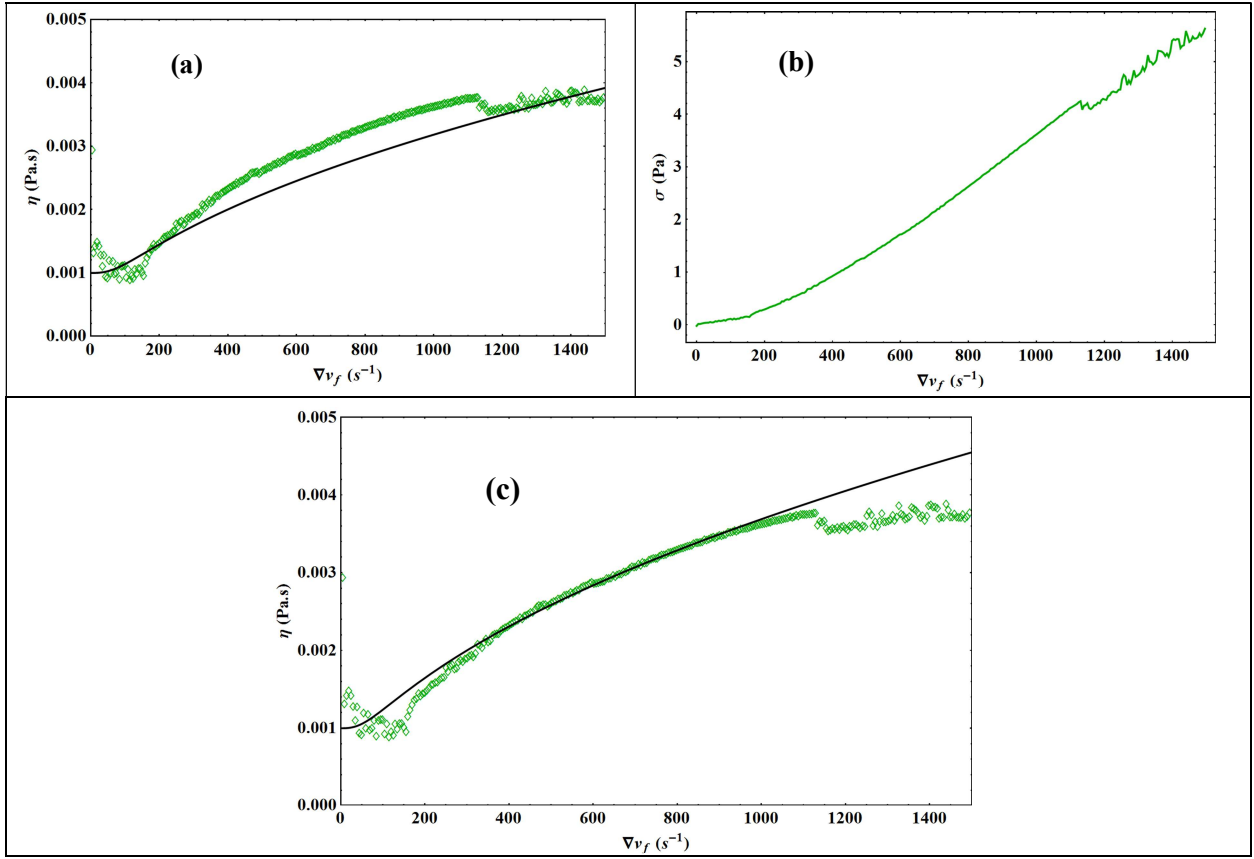

Fig. 13. Liquid water experiment at $T = 293.15$ K with the Couette cell by imposing a linear strain ramp. (a) Experimental results (green diamonds) versus theoretical model (black curve with $d_N = 1.6787$ cm and $C_K = 0.01086$) for the viscosity as a function of the velocity gradient. (b) The experimental stress versus the velocity gradient. (c) Experimental results (green diamonds) versus theoretical model (black curve with $d_N = 1.6787$ cm and $C_K = 0.00617$) for the viscosity as a function of the velocity gradient.

It may seem strange at first sight that the viscosity increases due to wall slip. However, this is only true in subregion 3 where the influence of the external disturbance can no longer be neglected. The explanation is as follows: a decrease in the value of $K$ induces an increase in the parameter $\lambda$ because of the decrease in the celerity $c_0 = \sqrt{K/\rho}$ (see Eq. (17)), and the decrease in $K$ in the expression of the liquid term of the viscosity $\eta_l = K\,\tau/H_N(v, \lambda)$, which is preponderant in the liquid phase, is more than compensated for by the decrease in $H_N(v, \lambda)$, thus an increase in the ratio. In other words, the slip effect is largely compensated by the increasing influence of the external perturbation introduced in the term $H_N(v, \lambda)$. Thus, the introduction of the parameter $\lambda$ provides a clear physical explanation of the shear-thickening phenomenon despite the presence or absence of wall slip.

The analysis of subregion 3 allowed us to determine the parameters $d_N$ and $C_K$ in each experiment. The analysis was done in relative form using an empirical value of the viscosity $\eta_0$. To determine an absolute value of $\eta_0$, the value of the dissipative distance $d$ must be fixed. To be consistent, the analysis of subregion 1 should be described using the inertial mode theory presented in Ref. 7. This theory introduces the notions of action temperature $T_A$, of viscous mass $K_{A0}^t$ (further details on this notion are given in appendix C) and of threshold stress $\sigma_T = \left(K_N + K_{gas}\right)\dfrac{\xi}{e}$, where $K_N = \dfrac{K}{H_N\left(v, \lambda = 0\right)}$ represents the macroscopic static shear elastic modulus, $K_{gas}$ the shear elastic modulus of the released gas and $\xi$ the correlation length between the fluid basic units. The parameter $e$ represents a characteristic distance of the experimental setup. In the case of the plane-plane rheometer with a small air-gap $e_g$, it was shown in Ref. 7 that for these conditions $e = d = e_g$. It has been shown that $\xi$ is compatible with the value $\xi = e_g$ at zero strain (corresponding to a zero-action temperature) and then decreases when the strain increases until reaching the value $\xi_0$ corresponding to an action temperature consistent with the establishment of a Newtonian regime. Let us recall that $\xi_0$ is an intrinsic property of the fluid which represents the distance over which the fluctuations of the unit cells are correlated in the bulk phase at thermodynamic equilibrium (see Ref. 6).

Eq. (28) of Ref. 7 allows us to define an effective viscosity $\eta_{eff}$ which must tend towards the known viscosity of the fluid $\eta_0$ when the action temperature $T_A$ becomes sufficiently large in front of the reaction temperature $T_{A0}$ so that we can consider the flow regime as Newtonian. Thus, by definition we write:

$$\eta_{eff} = C_{cal}\,\frac{E_A(X)\,d}{v_f(X,t)} \tag{23}$$

where $v_f(X,t)$ is determined by Eq. (28) of Ref. 7. By construction, Eq. (23) is such that $\eta_{eff}$ tends to $\eta_0$ when $T_A \gg T_{A0}$ and $t \gg \omega_c^{-1} = \tau_c$ where $\omega_c$ represents the cut-off pulsation of the inertial modes. In the case of the experiment with the HAAKE viscometer or with the Couette cell, $E_A(X)$ is identified with the experimentally determined or imposed stress $\sigma$ and $v_f(X,t)$ with the maximum radial velocity. The parameter $C_{cal}$ represents the calibration constant that must be applied to $\eta_0$ to compensate for the offset of the experimental data corresponding to the plateau of subregion 2.

As for subregion 3, we will start by analyzing the double cone experiments. The numerical values of the parameters needed for the modeling are grouped in Table 6.

| Name (unit) | Value |
|---|---|
| $K$ (GPa) | 2.95364 |
| $c_0$ (m/s) | 1720.33 |
| $K_N$ (Pa) | 18318.7 |
| $K_{gas}$ (Pa) | 652.8 |
| $d_N$ (cm) | 1.3144 |
| $\xi_0$ (Å) | 3.71768 |
| $v$ | 1.71492 |

Table 6. Numerical values of the fundamental parameters for liquid water at atmospheric pressure and 293.15 K when using the modeling from Ref. 6 and geometric characteristics of the double cone experiment.

Fig. 14a shows that Eq. (23) reproduces the data with a well-centered deviation. The deviation in subregion 2 is less than ±2% in agreement with the accuracy of the measurements. In subregion 1, the deviation increases up to ±4% but this deviation is only the result of the oscillations of the shear stress $\sigma$. Indeed, Fig. 14b shows that the oscillations of $\sigma$ are strong at the beginning of the experiment until a time $t \sim 20$ s. These oscillations are then reflected in the calculation of $\eta_{eff}$ in such a way that the value of the resulting deviation is consistent with these oscillations of $\sigma$. Therefore, Eq. (23) provides the best possible representation of the data.

Note that the parameters $d$ and $e$ fix the absolute values of viscosity while $\xi$ allows describing their variations in subregion 1. The values of these parameters are in accordance with what is expected for this type of experiment. Indeed, it is found that $d$ is slightly smaller than the "reference" value equal to 100 µm (see Ref. 6). Similarly, it is found that $e$ is slightly higher than the liquid thickness at distance $R$ corresponding to the rotor radius for the bottom cone. Since the upper double cone must be considered, it is normal to find a value of the order of 1 mm.

Let us focus on the variations of the effective viscosity in subregion 1. These variations are determined by the variations of the correlation length $\xi$. This is what was shown in Ref. 7 such that $\xi$ decreases as the strain increases. Similarly, $\xi$ should decrease when the strain rate increases. Given the boundary limits on $\xi$ previously mentioned, it turns out that this variation can be empirically described by the following general function:

$$\xi^*(\sigma) = \frac{\xi}{\xi_0} = 1 + \left( \frac{e}{\xi_0} - 1 \right) \frac{\exp\left( -\left( \sigma/\sigma_1 \right)^{\varepsilon_1} \right)}{1 + \left( \sigma/\sigma_2 \right)^{\varepsilon_2}} \tag{24}$$

where $\sigma_1$, $\sigma_2$, $\varepsilon_1$ and $\varepsilon_2$ are 4 empirical constants. It is immediately seen that Eq. (24) has the correct boundary properties when $\sigma = 0$ and $\sigma >> \sigma_1$. We preferred here a description in terms of the shear stress $\sigma$ rather than in terms of strain rate which allows to have a single input parameter for the whole modeling. The evolution of the parameter $\xi$ is shown in Fig. 14c and we note that the variation is limited to the subregion 1 at low strain rates. Therefore, as expected, the value of $\xi$ is equal to $\xi_0$ in subregion 2.

The knowledge of the parameter $\xi$ allows then to represent the variation of the reduced action temperature $T_A^* = T_A/T_{A0}$. It has been shown in Ref. 7 that the regime can be considered as Newtonian when $T_A^*$ reaches values of the order of 100. Fig. 14c shows that $T_A^*$ evolve between 25 and 35 in subregion 2. This mean that the liquid in this subregion 2 has not reached a perfectly Newtonian behavior and the measured viscosity has therefore not reached its Newtonian limit although it is numerically very close to it. Fig. 14c also shows that the value $T_A^* = 1$ is exceeded very quickly corresponding to very low strain rates (i.e. $\sim 10$ s$^{-1}$). Thus, the measurements made essentially correspond to a regime in which the medium behaves as a liquid.

Fig. 14d shows that the combination of Eqs. (19), (23) and (24) allows a consistent quantitative representation of the experimental data for all subregions. It can be seen that the Newtonian plateau simply corresponds to a flat minimum in the measured effective viscosity.

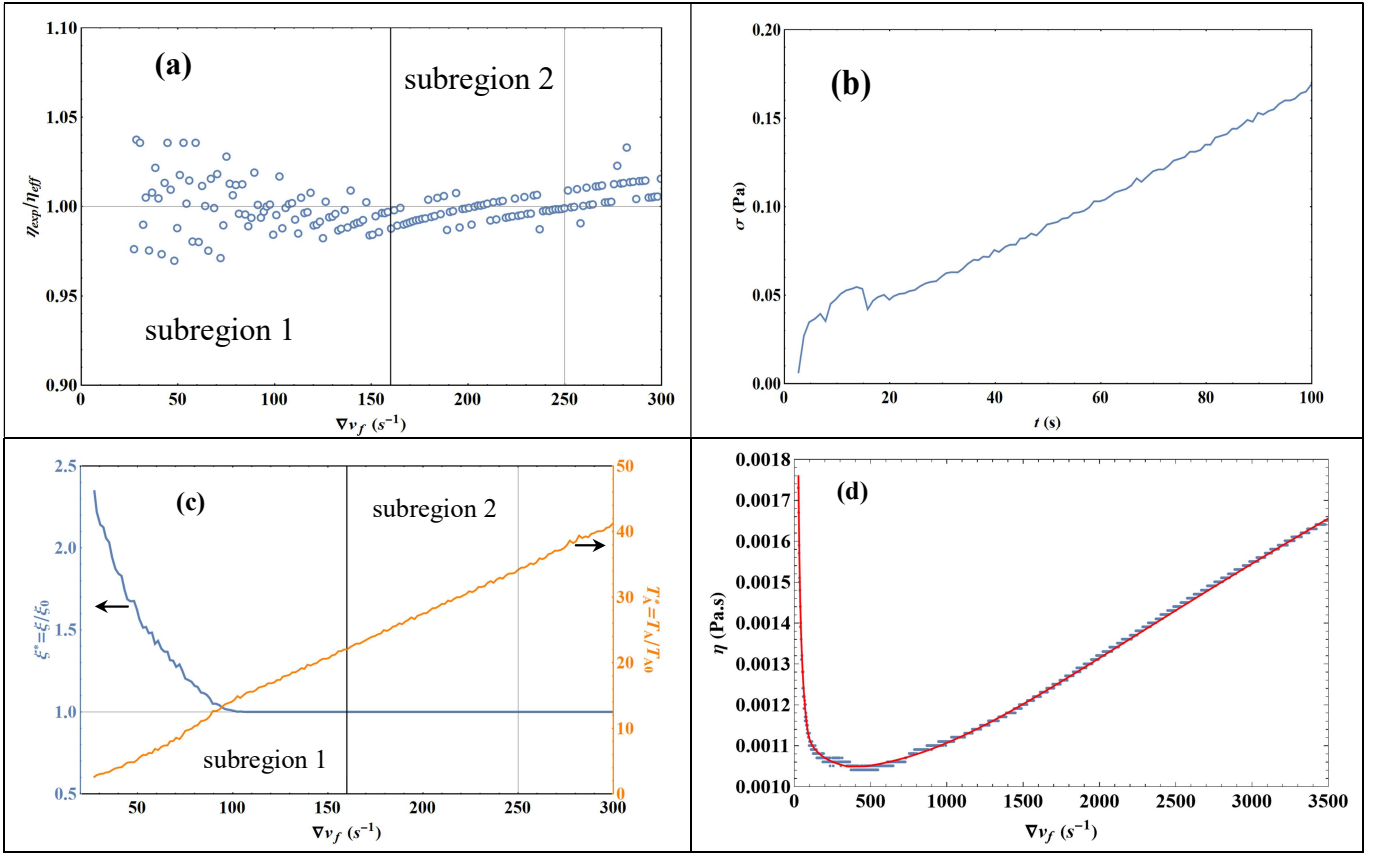

Fig. 14. Liquid water experiment at $T = 293.15$ K with the HAAKE double cone by imposing a linear strain ramp. The model parameters specific to this run are: $C_{cal} = 1.03789$, $d = 88.5$ µm and $e = 0.0908$ cm. (a) Deviation of the experimental results with Eq. (23). (b) Stress variation as a function of time $t$ for the first 100 seconds. (c) Representation of the reduced correlation length $\xi^*$ and the reduced action temperature $T_A^*$ as a function of velocity gradient using the experimental stress $\sigma$ as input parameter ($\sigma_1 = 0.099478$ Pa, $\sigma_2 = 1.17158 \times 10^{-4}$ Pa, $\varepsilon_1 = 10.358$ and $\varepsilon_2 = 2.402$). (d) Experimental results (blue points) versus theoretical model (red curve) for the viscosity as a function of the velocity gradient for all the subregions.

Above, an example was presented where a linear strain ramp was imposed. We recall here that the results when imposing a linear stress ramp give exactly the same results and therefore lead to the same analysis.

As for subregion 3, analysis of the data from the simple cone and Couette cell experiments requires replacing $c_0$ in Eq. (23) by the effective celerity $c_{eff} = \sqrt{K_{eff}/\rho}$ as previously explained.

We will analyze the results of the simple cone experiments. The numerical values of the parameters needed for the modeling are grouped in Table 7. It is observed that the decrease of the shear elastic constant $K$ leads to a strong decrease of the equilibrium macroscopic parameter $K_N = K/H_N(v, \lambda = 0)$.

| Name (unit) | Value |
|---|---|
| $K_{eff}$ (GPa) | 2.2741 |
| $c_{eff}$ (m/s) | 1509.37 |
| $K_N$ (Pa) | 1728.6 |
| $K_{gas}$ (Pa) | 652.8 |
| $d_N$ (cm) | 3.1781 |
| $\xi_0$ (Å) | 2.7029 |
| $\nu$ | 1.71492 |

Table 7. Numerical values of the fundamental parameters for liquid water at atmospheric pressure and 293.15 K when using the modeling from Ref. 6 and geometric characteristics of the simple cone experiment.

Although the parameters have significantly different numerical values compared to those of the double cone experiment, Fig. 15 shows that the analysis of the experiments with the simple cone is very similar to that of the double cone except that there is more noise in the determination of $\sigma(t)$ which is reflected in the calculation of $\eta_{eff}$. An interesting difference lies in the numerical value of the parameter $e$ which is much closer to the liquid thickness at distance $R$ corresponding to the rotor radius for the bottom cone. This result is consistent with what is expected since for these experiments the liquid height exceeds the thickness at distance $R$ corresponding to the rotor radius for the bottom cone by only 1 mm. The other difference that can be observed is that subregion 2 is shifted to lower values of the velocity gradient but we still have $\xi = \xi_0$. This is consistent with the fact that $T_A^*$ takes a higher value for the same velocity gradient in the simple cone experiments than with the double cone. It is observed that the medium reaches the same liquid-like regime in subregion 2 with $T_A^*$ values between 25 and 40.

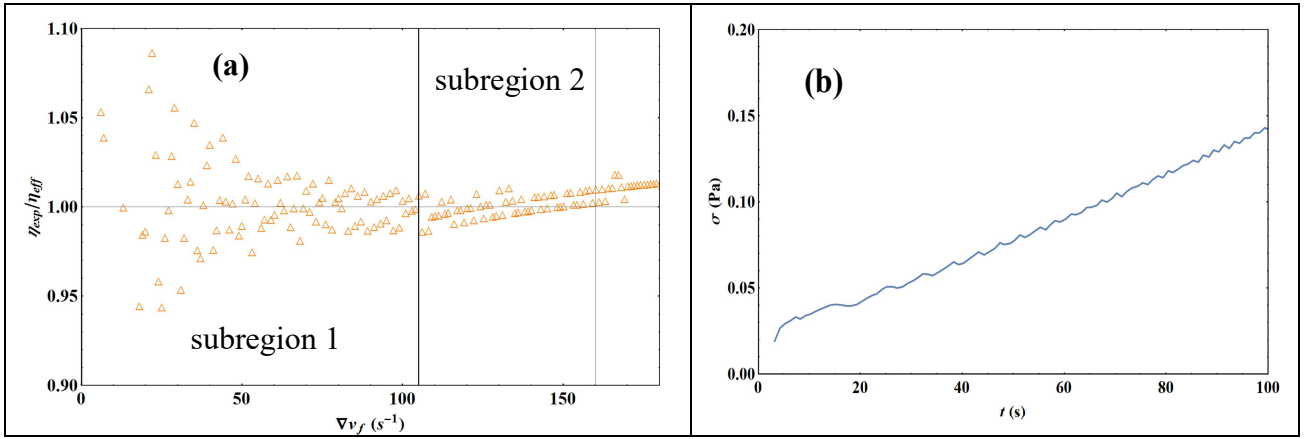

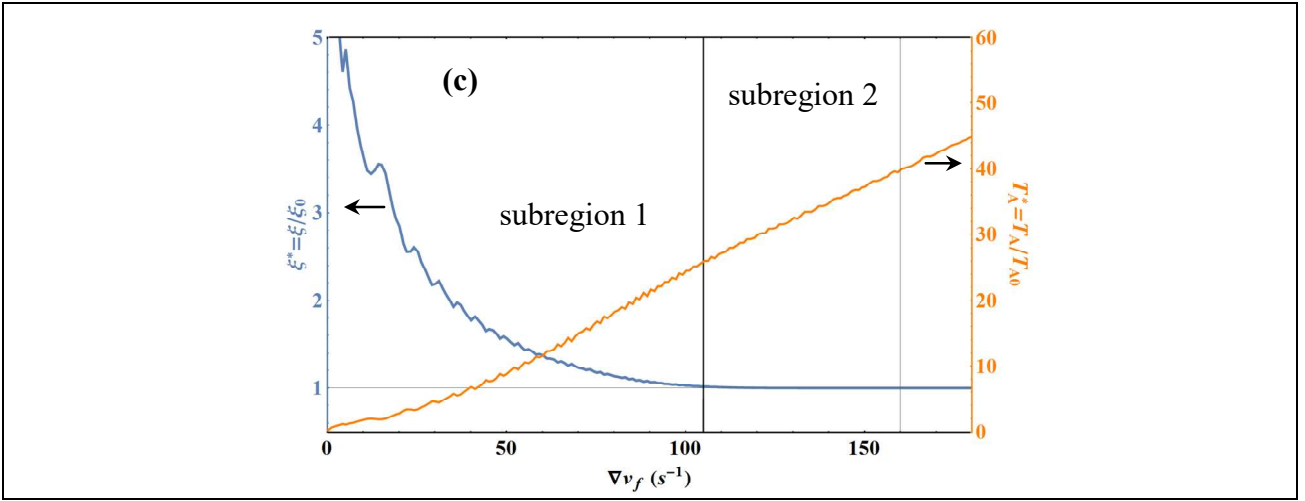

Fig. 15. Liquid water experiment at $T = 293.15$ K with the HAAKE simple cone by imposing a linear strain ramp. The model parameters specific to this run are: $C_{cal} = 1.42564$, $d = 110$ μm and $e = 0.0618$ cm. (a) Deviation of the experimental results with Eq. (23). (b) Stress variation as a function of time $t$ for the first 100 seconds. (c) Representation of the reduced correlation length $\xi^*$ and the reduced action temperature $T_A^*$ as a function of velocity gradient using the experimental stress $\sigma$ as input parameter ($\sigma_1 = 0.12813$ Pa, $\sigma_2 = 8.3848 \times 10^{-5}$ Pa, $\varepsilon_1 = 4.527$ and $\varepsilon_2 = 1.945$).

The experiments with the Couette cell lead to a very strong decrease of the effective elastic constant of the medium which leads to a rather strong decrease of most of the parameters as can be seen in Table 8. Thus, it is now the elastic constant of the released gas that dominates whereas the value of $K_{gas}$ was negligible in the experiment with the double cone and about three times lower in the experiment with the simple cone. This shows the importance of taking into account the gas released during the shear action.

| *Name* (unit) | *Value* |
|---|---|
| $K_{eff}$ (GPa) | 0.06675 |
| $c_{eff}$ (m/s) | 258.59 |
| $K_N$ (Pa) | 4.9179 |
| $K_{gas}$ (Pa) | 652.8 |
| $d_N$ (cm) | 1.6787 |
| $\xi_0$ (Å) | 1.2857 |
| $v$ | 1.71492 |

Table 8. Numerical values of the fundamental parameters for liquid water at atmospheric pressure and 293.15 K when using the modeling from Ref. 6 and geometric characteristics of the Couette cell experiment.

Fig. 16 shows that the results obtained with the Couette cell are quite similar to those obtained with the simple cone, which is consistent with the fact that these two experiments have in common the existence of a free surface. The deviation obtained in Fig. 16a is quite high but consistent with the strong oscillations of the curve $\sigma(t)$ as can be seen on Fig. 16b. The deviation being well centered, this shows that Eq. (23) allows to reproduce the data as well as possible.

The major difference is that from the first recorded points, we have $\xi = \xi_0$. This very fast variation of $\xi$ is consistent with the fact that very large values of $T_A^*$ are quickly obtained, as can be seen in Fig. 16c. This also explains that subregion 2 is reached for lower values of the velocity gradient than in the previous experiments. The decrease of the effective viscosity in subregion 1 when the strain rate increases is no longer related to the evolution of $\xi$ but corresponds to the transient regime. Indeed, in the experiments with the HAAKE viscometer, the characteristic time $\tau_c$ of this transient regime is of the order of 0.1 s, i.e. the first experimental point recorded already corresponds to at least $10\,\tau_c$. On the other hand, in the experiments with the Couette cell, the characteristic time $\tau_c$ is of the order of 10 s. Subregion 2 is reached after a time of about $4\,\tau_c$. Although the medium transits very quickly to a Newtonian liquid-like regime, there is still a time needed for the steady state to set in and thus for subregion 2 to be established. This transient regime was "hidden" by its very short duration in the HAAKE experiments.

It is also important to note that the value of $e$ is here exactly equal to the value of the Couette cell air-gap which is perfectly consistent with what is expected for this type of experiment.

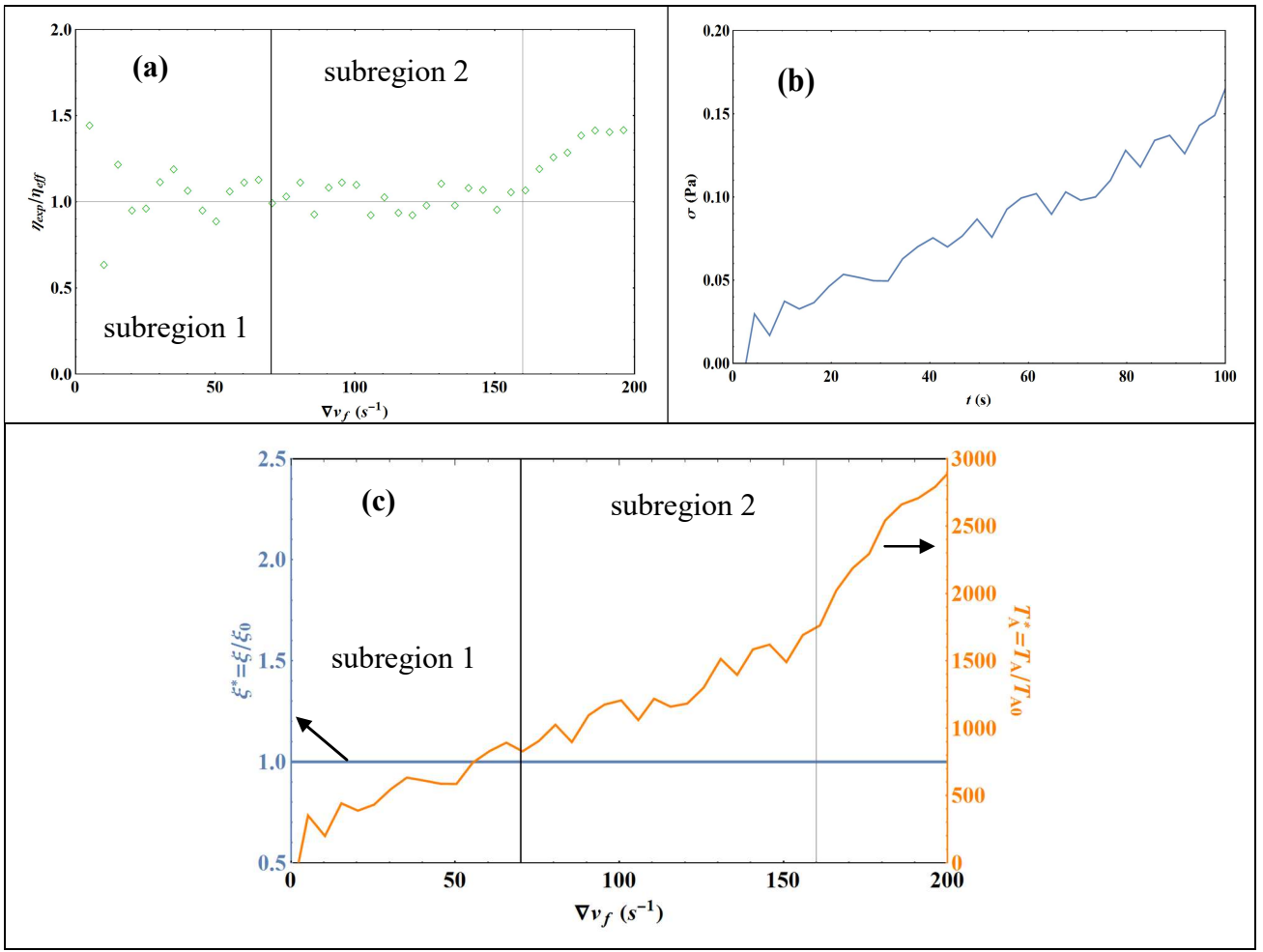

Fig. 16. Liquid water experiment at $T$ = 293.15 K with the Couette cell by imposing a linear strain ramp. The model parameters specific to this run are: $C_{cal}$ = 0.95707, $d$ = 320 μm and $e$ = 0.1 cm. (a) Deviation of the experimental results with Eq. (23). (b) Stress variation as a function of time $t$ for the first 100 seconds. (c) Representation of the reduced correlation length $\xi^*$ and the reduced action temperature $T_A^*$ as a function of velocity gradient using the experimental stress $\sigma$ as input parameter.

The set of experiments described in this section has allowed a wide range of parameter space to be explored, thus allowing a thorough test of the theoretical model. The experimental results shown in Fig. 9 find here a unique theoretical framework which allows to account for them and provides a coherent physical interpretation.

To conclude this section, Fig. 17 presents an example of results obtained with liquid n-octane. The strong resemblance with the results obtained with water suggests that the present model applies to all liquids, probably in their whole phase diagram. This is also in substance what Heyes (Ref. 4) wrote in his conclusion:

> "It was discovered here that simple liquids (e.g. argon, chlorine and water) behave rheologically [i.e. they should shear thin and shear thicken] in much the same way as these more chemically complicated fluid mixtures [e.g. mineral oil, polysaccharide xanthan gum]."

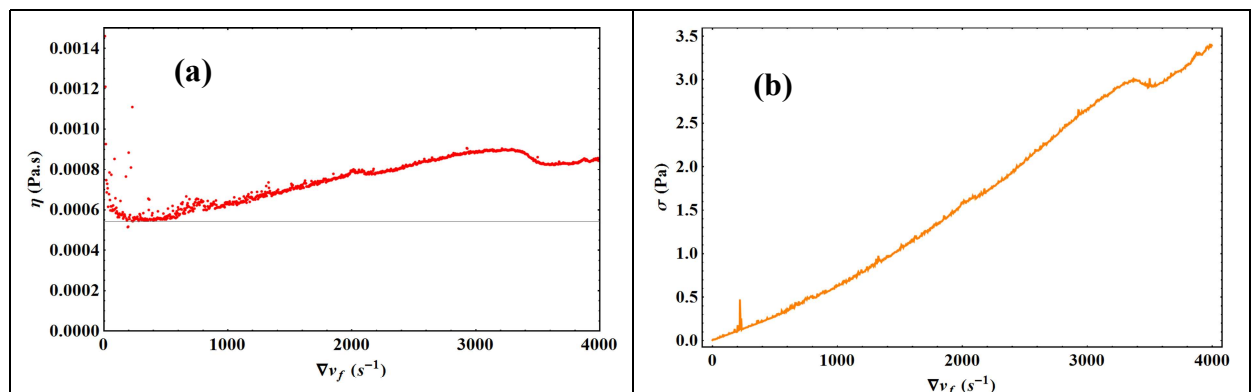

Fig. 17. Liquid n-octane experiment at atmospheric pressure and $T = 293.15$ K with the HAAKE simple cone by imposing a linear strain ramp. (a) Viscosity variation as a function of velocity gradient. The black horizontal line represents the expected viscosity value according to Ref. 18. (b) Stress variation as a function of velocity gradient.

## 4 Synopsis and conclusion

Contrary to standard (local) molecular models which are based on short-range interactions in space and on Markovian processes in time, thus without spatial nor temporal memory, the present (global) model is completely the opposite since it is based on long-range interactions in space, described by elastic modes (spatial memory), and hereditary processes in time, described by inertial modes (temporal memory), with the use of functionals introducing fractional derivatives. Let us recall that the fractional derivative of a continuous mathematical function is the convolution of this function with a power law, and it is this notion of convolution that introduces the memory aspects. The development of the model introducing two principles of equipartition: thermal energy (for space) and mechanical energy (for time) allows us to establish formal expressions for the measurable quantities (the observables) in terms of the model parameters. These parameters are of two types: intrinsic and extrinsic. **(i)** The intrinsic parameters are "necessary", they are of a physico-chemical nature and concern system at thermodynamic equilibrium: molecular composition, thermodynamic quantities such as the critical parameters, the phase diagram as well as all thermodynamic functions, to which are added, in the framework of the model, the elastic constant $K_0$, the shear celerity $c_0$, the cutoff wave-vector $q_c$ of the elastic modes, the correlation length $\xi_0$, the number of atoms/molecules in the basic unit $n_B$, all related to the thermodynamic parameters. **(ii)** The extrinsic parameters are "contingent": they depend on the type of experiment (here a "flow" in the broad sense of the term) that we want to carry out: the size, shape and volume of the samples through the fluctuative distance $d_N$ and dissipative distance $d$, the nature of the walls of the container through the parameter $K_N$, the parameter $\lambda$ associated with the average velocity gradient caused by the imposed mechanical stress, and possibly other additional external fields such as an electric or magnetic field for more complex experimental situations.

In Refs. 6 to 8, the model was limited to physical situations where the perturbation on the system is sufficiently small so that the theory could be developed assuming that the system remains at thermodynamic equilibrium during the measurement process. In the present paper, this perturbation is explicitly introduced in the model by adding an external energy term in the expression of the elastic energy functional which describes the coupling between the random thermal motions and the deterministic displacement induced by the mechanical perturbation.

It has been shown that the combination of this generalized elastic mode theory with the inertial mode theory developed in Ref. 7 allows to understand and describe to experimental precision, the results of a number of rheology experiments on liquid water in a broad range of experimental conditions. This demonstrates that the spatiotemporal aspects represented by these two theories are inseparable to analyze rheological experimental results.

The new results on liquid water presented here show non-Newtonian behavior except in a range of velocity gradients corresponding to the Newtonian plateau. For this limited range, which leads to viscosity values compatible with a Newtonian flow regime, it has been shown that the simplified form of the elastic mode theory as presented in Ref. 6 is sufficient to analyze the usual viscometry experiments.

The "shear-thinning" behavior observed at very low velocity gradients is explained by the dynamic phase transition of the inertial mode theory from a solid-like regime to a liquid-like regime, such that the correlation length $\xi$ introduced in the model decreases from a characteristic distance $e$ of the experimental set-up, allowing to define the threshold stress, to the correlation length $\xi_0$ defined by Eq. (10) of Ref. 6. This behavior is in perfect agreement with the fact that sufficiently close to thermodynamic equilibrium (not net flow), any finite volume of fluid must be considered as a solid, as abundantly demonstrated by numerous rheology experiments with sub-millimeter size samples at very low shear stress or strain amplitude and frequency (see Ref. 7 for more details).

When the velocity gradient is increased further beyond the Newtonian plateau, the experimental results show that the liquid water viscosity increases. This "shear-thickening" behavior is explained by the fact that the influence of the external energy, characterized in the model by the non-dimensional parameter $\lambda$, can no more be neglected and may become preponderant at sufficiently high velocity gradients. In other words, numerically, one can no longer neglect the disturbance introduced by the measurement process, and this aspect is taken up again in Appendix A in connection with the measurement problem in Quantum Mechanics. Thus in the present modeling, this shear-thickening phenomenon which becomes measurable at sufficiently high velocity gradients is associated with the increased importance of the external energy injected in the system which tends to decrease the amplitude of the thermal fluctuations, thus increasing the effective viscosity and the associated shear elasticity. For even larger stresses, one reaches the turbulence domain discussed in Appendix B and, at very high stresses corresponding to "high energy" situations, the size of the thermal clouds tends towards zero so that the particles can now be considered as conventional material points with zero size which can be localized in space, and whose displacements are described by the deterministic equations of classical mechanics.

This model, which has allowed to represent within experimental error the data in an ideal case were the no-slip condition is well satisfied, can also be applied in more complex cases where slip-phenomena occur, by slightly modifying the model via the introduction of an effective shear elastic constant $K_{\text{eff}}$.

It has also been mentioned that the same analysis can be made with other liquids suggesting that this modeling can be extended to phenomena implying a large variety of fluids and probably in other domains of physics.

A very important feature of this model is that, as soon as a mechanical action, however small, is made on a system, more generally as soon as a system is out-of-thermodynamic equilibrium, all the integrals involved in the expressions of the experimental quantities never diverge whatever the finite or infinite size of the system. This is inherent to any measurement, since one cannot imagine obtaining any information on a system without exchanging energy with it, thus without perturbing it. This feature of the theory is to be compared with the problem of measurement in quantum mechanics, but which here applies whatever the scale of the system.

In conclusion, in all practical situations, the present modeling has shown that the random aspects associated with thermodynamics, and the deterministic aspects associated with mechanics coexist, and one or the other is predominant only in extreme physical situations. The random aspects dominate as long as the mechanical energy is very low in front of the thermal energy, and *vice-versa*. But this is only true numerically, at the experimental level. In all rigor, the two aspects are inseparable for the description of the real world. This combination of the random and the deterministic aspects proposed by the present modeling is probably one key to solve the controversial and long-standing problem of the interpretation of Quantum Mechanics.

Some examples of application of this theory extended to different domains of physics will be presented in forthcoming papers.

## 5 ACKNOWLEDGEMENTS

We would particularly like to thank Drs. Pierre Terech and Olivier Diat (CEA Grenoble, France) for their help with the experiments made with the double cone rheometer and the Couette rheometer, respectively.

We thank Dr. L. Noirez from the Léon Brillouin laboratory (CEA, France) for many helpful suggestions and discussions.

This work benefited from the support of the project ZEROUATE under Grant ANR-19-CE24-0013 operated by the French National Research Agency (ANR).

## 6 APPENDIX A: "Uncertainty Relationship"

Considering Eqs. (17) and (13), $<\left|u^2\right|>$ can be rewritten in a form that is reminiscent of the uncertainty principle of Quantum Mechanics (QM). First of all, let's remember that the product $\left(\nabla v_f\right)_{\mathrm{av}} l_{pm}$ can be interpreted as an average relative velocity $\bar{v}_m$ between two neighboring atoms (or molecules) of the medium. It comes then:

$$<\left|u^2\right|> \bar{v}_m^2 = \frac{\lambda^2 H_N(v,\lambda)}{\pi^2}\frac{k_B T\, q_c}{\rho} \tag{A-1}$$

By introducing the volume per atom (or per molecule) $\mathcal{V}_{pm} = m/\rho$ where $m$ represents the mass of an atom (or of a molecule) which is related to the Compton wave-vector $q_{\mathrm{Comp}} = mc/\hbar$ ($c$ being the celerity of light and $\hbar$ the reduced Planck constant) associated with this atom (or molecule), the following expression can be deduced:

$$< \left| u^2 \right| > \left( m \bar{v}_m \right)^2 = \left( \lambda^2 H_N \left( v, \lambda \right) \; \Phi_{\text{eth}} \right) \frac{\hbar^2}{4} \qquad \text{(A-2)}$$

where $\Phi_{\text{eth}} = \dfrac{v_{pm}}{\dfrac{\Lambda_c}{4} \dfrac{\Lambda_{\text{Comp}}}{4} \dfrac{l_{\text{eth}}}{2}}$ is a dimensionless coefficient that depends only on temperature and density of the medium. The parameters $\Lambda_c = 2\pi / q_c$ and $\Lambda_{\text{Comp}} = 2\pi / q_{\text{Comp}}$ are the elastic mode cut-off wavelength and Compton wavelength, respectively. The quantity $l_{\text{eth}} = \dfrac{2\hbar c}{k_B T}$ represents a characteristic distance whose physical meaning will be discussed in a future paper.

We thus see appearing in the right-hand side of Eq. (A-2) a dimensionless coefficient in the parenthesis which depends on the temperature and the density of the medium, and on the action bringing the out-of-equilibrium of the medium through the parameter $\lambda$: for a given temperature and density, $\Phi_{\text{eth}}$ is a constant such that the coefficient in parentheses in the right-hand member of Eq. (A-2) varies only as a function of $\lambda$; and this coefficient decreases when $\lambda$ increases. Thus, the greater the external action on the medium, the smaller $< \left| u^2 \right| >$ is and therefore the more localized the basic unit is in space in accordance with what is expected from the wave mechanics.

It can be interesting to ask for which value of $\lambda$ the coefficient of $\hbar^2 / 4$ reach the unit value. In a general way it is necessary to solve the equation $\lambda^2 H_N \left( v, \lambda \right) = \Phi_{\text{eth}}^{-1}$. But the values of $\lambda$ which satisfy this equation are large in front of 1 and therefore one can consider the asymptotic limit of $H_N \left( v, \lambda \right) \approx \left( 3 \lambda^{1+v} \right)^{-1}$ to the first approximation (see Eq. (9) and Eq. (13)). It is then deduced that $\lambda$ which satisfies the sought condition is simply written:

$$\lambda_{MQ} = \left( \frac{\Phi_{\text{eth}}}{3} \right)^{\frac{1}{v-1}} \qquad \text{(A-3)}$$

Eq. (A-3) is a function of $T$ and $\rho$. Fig. 18 shows in the case of liquid water in normal and supercooled phases the function $\lambda_{MQ}(T, \rho)$. We first notice that the values of $\lambda_{MQ}$ are large in front of unity except near 200 K. It appears that in the normal liquid phase the variations are small so that $\lambda_{MQ} \approx 500$ while in the supercooled phase the variations are larger with a peak around 250 K. Any value of $\lambda$ that is greater than $\lambda_{MQ}(T, \rho)$ leads to a coefficient in front of $\hbar^2 / 4$ that is smaller than 1 and vice versa.

If we now transform the function $\lambda_{MQ}(T, \rho)$ in terms of $\left( \nabla v_f \right)_{\text{av, MQ}}(T, \rho)$ then it is obtained that the variations are much smaller and uniform such that $\left( \nabla v_f \right)_{\text{av, MQ}}$ varies between $10^{14}$ s$^{-1}$ at 200 K and $10^{15}$ s$^{-1}$ in the normal liquid phase. These gradient values are very large and far exceed the values of the experiments we have analyzed in this paper. They are also large in front of the values of $\lambda$ for which turbulence appears in the usual experiments as described in Appendix B. Therefore, the coefficient in front of $\hbar^2 / 4$ is greater than 1 in the usual experiments with liquid water.

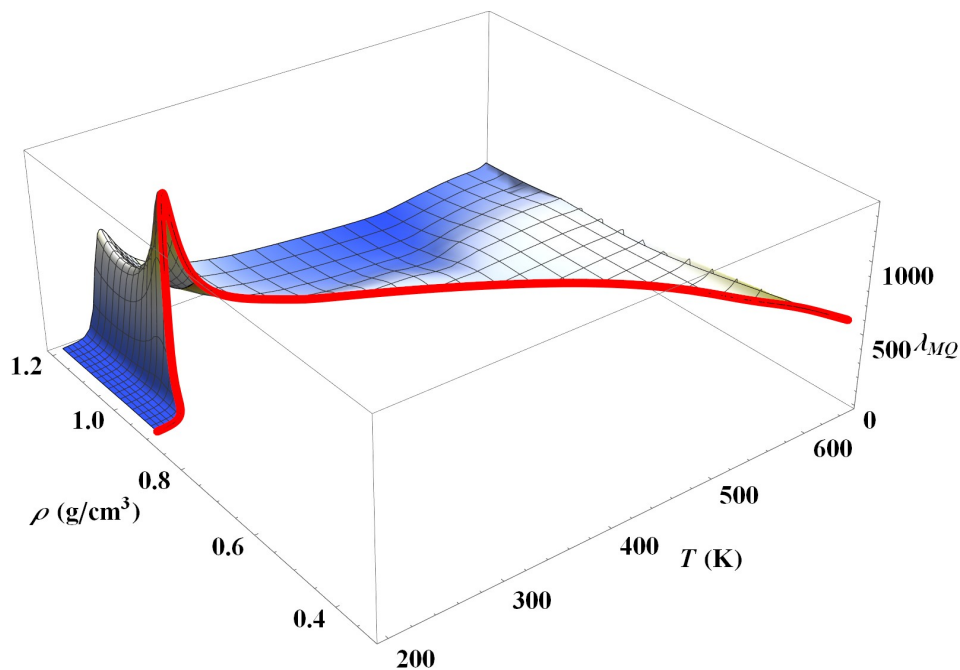

Fig. 18. Representation of Eq. (A-3) in the normal liquid and supercooled phases of water between 200 K and the critical temperature. The thick red curve represents the liquid coexistence curve in the normal liquid phase and then extends into the supercooled phase with the atmospheric isobar.

Fig. 19 shows what Eq. (A-3) gives in the gaseous phase of water. It can be seen that $\lambda_{MQ}(T,\rho)$ now depends strongly on the density but almost not on the temperature. Although the numerical values of $\lambda_{MQ}$ here are large, their correspondences in terms of $\left(\nabla v_f\right)_{\mathrm{av,\,MQ}}$ show little variation between $10^{15}$ s$^{-1}$ and $10^{16}$ s$^{-1}$. These values are thus comparable to those obtained for the liquid phase. These values can therefore be reached more easily with gases than with liquids.

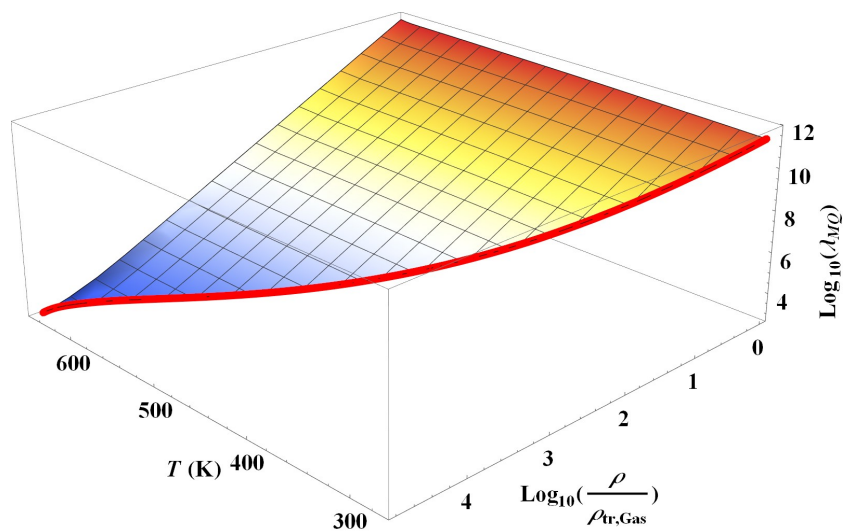

Fig. 19. Representation of Eq. (A-3) in the gaseous phase of water between the triple point and critical temperature, and from the triple point gas density to the critical one. The thick red curve represents the vapour coexistence curve.

In conclusion of this Appendix, it appears that the present modeling has a certain link with QM through the notion of uncertainty. Indeed, in both models, there is a source of fundamental indeterminacy through the impossibility of defining the position and velocity of a particle in the deterministic sense of classical mechanics.

In the present modeling, the particles necessarily belong to a fluid, and the indeterminacy on the position is related to the existence of the thermal cloud whose size is related to the experimental conditions, whereas in QM, an isolated particle exists *a priori*, but the indeterminacy is, according to the possible interpretations, either an intrinsic property or related to the problem of the measurement.

For the velocity, things are also different. In the present model, the velocity concerns the fluid and the indeterminacy is related to the difference in velocity between two neighboring particles caused by the existence of an average velocity gradient. The corresponding uncertainty principle follows from the fact that the theory predicts that the size of the thermal cloud is as an inverse function of the mean velocity gradient, which leads to Eq. (A-2). For QM, the notion of velocity exists per se, but the uncertainty is as for position, either an intrinsic property or related to the measurement problem as well. For QM, velocity and position are defined with respect to a fixed reference frame, whereas in the present model, these are defined with respect to the fluid in which there is fluctuation (because of the thermal cloud) and dissipation (because of the velocity gradient), the notion of laboratory reference frame not being directly involved, only to define the details of the experimental setup. All these notions and differences between the two descriptions will be clarified in a forthcoming paper concerning the effusion of a dilute-gas in "vacuum".

## 7 APPENDIX B: Criterion for transition to turbulence

In this appendix, we will see how one can define a criterion for transition to turbulence in the framework of the present modeling.

We have seen that the parameter $\lambda$ can be written as the ratio of two rates, the average velocity gradient $\left(\nabla v_f\right)_{av}$ and the inverse of the time $\tau_{mol} = l_{pm}/c_0$ associated with the characteristic intermolecular or interatomic distance. It was thus noticed that $\lambda$ has a microscopic meaning. Indeed, when $\lambda = 1$, it means that the velocity gradient becomes comparable to the response time of the reaction of the system at the microscopic level. For stronger gradients, then the molecules or atoms will behave as if they were isolated from each other. Indeed when $\lambda = 1$, we observe that $H_N(v, \lambda)$ is always close to its limit value given by Eq. (9) whatever the value of $N$ and $v > 1$. Given the typical values of $\tau_{mol}$ which is on the order of picoseconds or less, the velocity gradient values that correspond to the condition $\lambda = 1$ are of the order of terahertz.

The transition to turbulence corresponds to a macroscopic response of the system and not a microscopic one. However, the response time of the system at the macroscopic scale is given by the time $\tau$ and this time is related to the time $\tau_{mol}$ such that: $\tau = \dfrac{d}{l_{pm}}\tau_{mol}$. By definition, the time $\tau$ reflects the natural "capacity" of the macroscopic system to dissipate the energy (finally transformed into heat) which is communicated to it during an action made on this system. Depending on whether the velocity gradient is smaller or larger than $\tau^{-1}$, the system is expected to respond differently, i.e. it is around $\left(\nabla v_f\right)_{av}\tau = 1$ that a qualitative change in the system's response to the action producing the velocity gradient will occur. The time $\tau$ being typically of the order of $10^{-8}$s, it is deduced that the condition will be fulfilled when the velocity gradient will be of the order of ten megahertz. In section 2.2, the following notation has been introduced:

$$\Lambda = \left(\nabla v_f\right)_{\mathrm{av}} \tau \tag{B-1}$$

Given Eq. (17), the following relation is obtained:

$$\lambda = \Lambda \frac{\tau_{\mathrm{mol}}}{\tau} = \Lambda \frac{l_{pm}}{d} \tag{B-2}$$

Therefore, the condition $\Lambda = 1$ will occur for a smaller value of the velocity gradient than for the condition $\lambda = 1$. Considering Eq. (B-2), it is deduced that the typical order of magnitude of $\lambda$ when $\Lambda = 1$ is $\lambda \approx 10^{-6}$.

It can be noticed that the dimensionless number $\Lambda$ resembles a Weissenberg number except that the time involved here is not the stress relaxation time but the time for the shear information to propagate over the distance $d$. Given Eq. (B-1), we determine the expression for the average power dissipated per unit volume such that:

$$\left(\mathscr{P}_\eta\right)_{\mathrm{av}} = \eta \left(\nabla v_f\right)_{\mathrm{av}}^2 = \frac{\Lambda^2\, K_{\mathrm{A0}}^{\mathbf{t}}\, c_0^2}{\mathcal{V}_B\, \tau} \tag{B-3}$$

where $K_{\mathrm{A0}}^{\mathbf{t}}$ represents the viscous mass (see appendix C) and $\mathcal{V}_B$ the volume per basic unit. The numerator in Eq. (B-3) thus represents the average energy $\left(\mathscr{E}_\eta\right)_{\mathrm{av}}$ dissipated in the basic unit volume during the time $\tau$. At the turbulent transition, the average dissipated energy per basic unit is then simply written $\left(\mathscr{E}_\eta\right)_{\mathrm{av}} = K_{\mathrm{A0}}^{\mathbf{t}}\, c_0^2$ which recalls a famous formula of the special relativity.

In the liquid phase where the approximation $\eta \approx \eta_l$ can be considered, Eq. (15) allows to relate the parameter $\Lambda$ with the Reynolds number Re such that:

$$\mathrm{Re} \cong \frac{L}{d} \Lambda\, H_N\left(v, \lambda\right) \tag{B-4}$$

Given Eq. (B-4), the condition $\Lambda = 1$ results in:

$$\mathrm{Re}_{\mathrm{crit}} \cong \frac{L}{d} H_N\left(v, \lambda = \frac{l_{pm}}{d}\right) \tag{B-5}$$

Fig. 20 shows the values of the critical function $H_N\left(v, \lambda = \dfrac{l_{pm}}{d}\right)$ in the case of liquid water along the atmospheric isobar. It can be seen that this critical function increases with increasing temperature which means that the value of the critical Reynolds number increases, all other things being equal. This is consistent with the fact that the dynamic as well as the kinematic viscosities decrease with increasing temperature along the atmospheric isobar: the liquid being less viscous, the transition to turbulence is delayed.

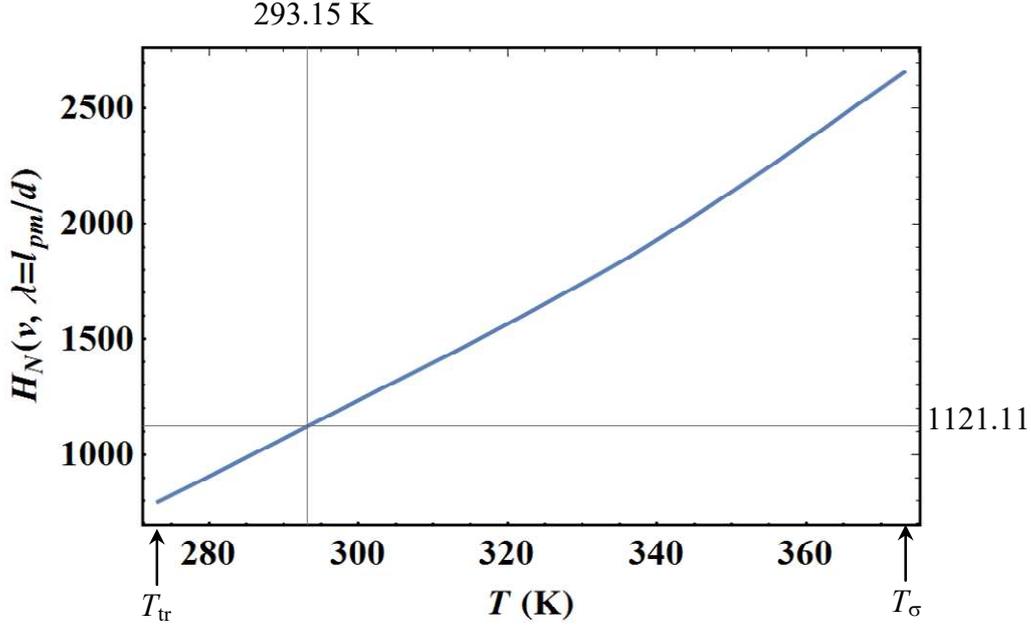

Fig. 20. Evolution of the critical function $H_N\left(v, \lambda = l_{pm}/d\right)$ (with $d = 0.01$ cm) versus the temperature along the atmospheric isobar of liquid water, between the triple point temperature $T_{tr}$ and the saturated vapor pressure curve temperature $T_\sigma$.

It is known that the value of the critical Reynolds number depends on the geometry of the experiment. Let us consider three well known cases of simple geometry:

- In the case of tubes, $L$ corresponds to the diameter of the tube $D$ and if the capillary tube is small enough then $d = D/2$ and the critical Reynolds number is such that $\text{Re}_{\text{crit}} \cong 2\,H_N\left(v, \lambda = \dfrac{l_{pm}}{d}\right)$. At room temperature, we deduce from Fig. 20 that $\text{Re}_{\text{crit}} \cong 2242$. This value typically corresponds to the minimum value allowed for this type of experimental device.

- In the case of channels, $L$ corresponds to the distance between the two parallel planes and if the thickness of the channel is small enough then $d = L$ and the critical Reynolds number is such that $\text{Re}_{\text{crit}} \cong H_N\left(v, \lambda = \dfrac{l_{pm}}{d}\right)$. At room temperature, we deduce from Fig. 20 that $\text{Re}_{\text{crit}} \cong 1121$. This value typically corresponds to the minimum value allowed for this type of experimental device.

- In the case of flow over a flat plate, $L$ corresponds to the critical distance $x_c$ to the upstream edge of the plate. Typical values of $x_c$ are of the order of magnitude of a few centimeters, therefore at room temperature the minimum value of the critical Reynolds number is $\text{Re}_{\text{crit}} \cong 112\,100$. This value typically corresponds once again to the minimum value allowed for this type of experimental device.

We deduce that the condition $\Lambda = 1$ is a very general condition that characterizes the fluid with respect to the translational turbulence instability in a given geometry.

This provides a picture and scenario of the transition to turbulence in the framework of the present modeling. Indeed, let us consider a region of the fluid of average dimension $L$ which undergoes a velocity gradient such that $\Lambda$ calculated with this distance $L$ is greater than 1. Then this region will divide into smaller fragments of dimensions $L' < L$ such that $\Lambda' = 1$, so as to evacuate sufficiently quickly the heat introduced by the gradient at the interfaces thus created. If there is a broad distribution of sufficiently strong gradients in the sample, then fragments will be formed whose size distribution will reflect the distribution of these gradients.

It is important to point out that this description should not be confused with that of Kolmogorov (Ref. 19). Indeed, we do not have here a cascade of scales within large eddies to smaller eddies but we can have a distribution of fragmented regions defined by $\Lambda = 1$ up to $\lambda = 1$ depending on the local velocity gradient.

In conclusion of this appendix, it appears that the model presented in all our previous papers as well as this one can be considered as a new approach to quantitatively describe fluid physics, not only in a range of relatively slow flow regimes, such as the solid shear-thinning regime, laminar regimes (Newtonian and shear-thickening), but also to predict the existence of dynamic transitions leading to chaotic regimes such as turbulence. It would be appropriate to try to describe the experimental results in these areas for a quantitative test of this model and to compare them with the analyses made by the standard theories of turbulence and chaos. This is the subject of future work.

## 8 APPENDIX C: Notion of viscous mass

In Ref. 7, the parameter $K_{A0}^{t}$ which has the dimension of a mass was introduced by analogy with the elastic constant $K$ of the elastic mode theory (Ref. 6). As mentioned, $K_{A0}^{t}$ is a global quantity associated with dissipative effects so that we will name it ***viscous mass***. The expression for $K_{A0}^{t}$ defined by Eq. (20) in Ref. 7 involves the function $H_N(v)$, which means that $K_{A0}^{t}$ was assumed to be independent of the dynamic state of the fluid. We have seen that in practice this approximation is sufficient in the Newtonian regime, but beyond this regime, the dependence on the parameter $\lambda$ can no longer be ignored and must be considered by replacing $H_N(v)$ by $H_N(v, \lambda)$. In the framework of the generalized formalism introduced in this article, it is then interesting to discuss more deeply this notion of viscous mass. Its expression is defined by Eq. (20) of Ref. 7 as the sum of two terms such that:

$$K_{A0}^{t} = m_{vi}^{l} + m_{vi}^{Knu} \tag{C-1}$$

where

$$m_{vi}^{l} = \frac{m_B}{H_N(v, \lambda)} \tag{C-2}$$

and

$$m_{vi}^{Knu} = m_B \frac{\rho_{Knu}}{\rho} \frac{\delta}{d} \frac{\sqrt{R_g T / M}}{c_0} \tag{C-3}$$

Both terms are scaled by the basic unit rest mass $m_B$ which is a local quantity.

The parameter $\rho_{Knu}$ represents the density of the released gas due to the shear stresses which induces the flow and $\delta$ is a characteristic distance which in the case of a Poiseuille type flow can be identified with the dissipative distance $d$.

Given the definition of the liquid and gaseous components $\eta_l$ and $\eta_{Knu}$ in the expression of the dynamic viscosity defined by Eqs. (13) and (16) in Ref. 6, one can deduce a rewriting in terms of the expressions of the viscous masses such that:

$$\eta_l = \frac{m_{vi}^l}{\mathcal{V}_B} c_0 \, d \tag{C-4}$$

and

$$\eta_{Knu} = \frac{m_{vi}^{Knu}}{\mathcal{V}_B} c_0 \, d \tag{C-5}$$

where $\mathcal{V}_B$ represents the volume per basic unit.

It is seen that Eqs. (C-4) and (C-5) can be written analogously in a form reminiscent of the expression from the kinetic theory of gases, but where the average velocity of the molecules/atoms is replaced by the shear celerity $c_0$, the mean free path by the dissipative distance $d$ and the mass of an atom/molecule by the viscous mass. In other words, the microscopic quantities of the kinetic theory of gases are replaced by macroscopic quantities at the sample scale. Despite a form of dissimilarity in the expression of the viscous masses, Eqs. (C-4) and (C-5) show a uniformity in the description of the viscous terms.

By inverting Eqs. (C-4) and (C-5) and introducing the characteristic time $\tau = d/c_0$, we derive the following general relationship:

$$m_{vi}^{l/Knu} = \eta_{l/Knu} \, \tau \, \frac{\mathcal{V}_B}{d^2} \tag{C-6}$$

which shows that the viscous mass is proportional to the viscosity, the characteristic time $\tau$ as well as to a length scale which depends on the geometry of the experiment.

The dependence of the mass on the dynamic state of the system, characterized by the parameter $\lambda$, which is equal to the ratio of two velocities, has some relation with the Special Relativity (SR) where the mass also depends on the ratio of two velocities. In both cases, it is an increasing function of this ratio. But the analogy stops there. In SR, the variation of the mass concerns the total energy of the moving system in a model without dissipation, whereas in the present modeling, it concerns the dissipation only. The other important difference is that, in SR, the velocity $v_f$ is that of the particle with respect to the laboratory reference frame and the reference velocity is the speed of light $c$, whereas in the present modeling, the former is the average relative velocity between two neighboring particles in the fluid $\bar{v}_m = \left( \nabla v_f \right)_{av} l_{pm}$, and the latter is the shear velocity $c_0$. In a normal fluid flow, the relativistic aspect is totally negligible as far as the transported energy is concerned, and therefore the total mass can be assimilated to the rest (or proper) mass, while the viscous "relativistic" aspect, although negligible in the Newtonian regime, becomes fundamental beyond, since it governs not only the variation of dissipation associated with the increase of viscosity (see Fig. 8), but also the appearance of instabilities such as turbulence (see appendix B).

Considering the analogy that we have pointed out with SR, it is interesting to push further the analysis by observing the evolution of the mass ratio $m_{vi}^{l/Knu}/m_B$. We will start with the analysis of the gaseous ratio because it does not depend on the dynamical state of the system (i.e. it does not depend on $\lambda$) but only on the existence of a shear stress which induces the flow. In Eq. (C-3), the quantity $\sqrt{R_g T/M}$ is identified with the square root of the variance $\left\langle v_M^2 \right\rangle$ of the Maxwell-Boltzmann Gaussian velocity distribution in the kinetic theory of gases. For the sake of simplification, we will assume here that $\delta = d$ which is generally the case in usual rheology experiments. Thus, the sought ratio for the released gas can be simply written:

$$\frac{m_{vi}^{Knu}}{m_B} = \frac{\rho_{Knu}}{\rho} \sqrt{\frac{\left\langle v_M^2 \right\rangle}{c_0^2}} \qquad (C-7)$$

In usual experiments, it has been shown in the case of water (Ref. 6), potassium and thallium (Ref. 8), that $\rho_{Knu} << \rho$ and $\left\langle v_M^2 \right\rangle \approx c_0^2$, therefore Eq. (C-7) implies that $m_{vi}^{Knu} << m_B$. Let us examine the conditions to obtain $m_{vi}^{Knu}/m_B = 1$. It was shown in Ref. 6 that the condition $\rho_{Knu}(\rho, T) = \rho$ leads to a density value $\rho_{g0}(T)$ that is much smaller than that of the gas at the triple point of water along the isotherm at 650 K (see Fig. 68 of Ref. 6). Thus, this situation occurs for dilute gases at high temperature. It is deduced that $\left\langle v_M^2 \right\rangle = c_0^2$ can also be written by using the perfect gas equation of state $\dfrac{P}{\rho} = \dfrac{c_{K0} K_0}{\rho_c} \left( \dfrac{\rho}{\rho_c} \right)^2$ (see section III.A of Ref. 6), where $P$ represents the hydrostatic pressure. By combining this result with the value of $\rho_{g0}$, it is deduced a value of pressure that satisfies $m_{vi}^{Knu}/m_B = 1$ along the considered isotherm such that $P = c_{K0} K_0 \left( \dfrac{\rho_{g0}}{\rho_c} \right)^3$. If we take the example given above from Fig. 68 of Ref. 6, the condition $m_{vi}^{Knu}/m_B = 1$ leads to a pressure value $P = 1.162 \times 10^{-11}$ Pa which is much lower than the value of the triple point pressure of water, a situation which is not reached in the usual experiments.

Let us now analyze the case of the liquid ratio $m_{vi}^l/m_B = 1/H_N(v, \lambda)$. The description of the function $H_N(v, \lambda)$ is done in section 2.1 but it appears here as a physical quantity relating the rest mass of a basic unit to the viscous mass. Let us first note that $m_{vi}^l/m_B \cong 1$ when $v = 0$ and $\lambda = 0$ for $N$ sufficiently large (i.e. $H_N(v = 0, \lambda = 0) = 1 - \dfrac{1}{N}$), in other words, the viscous mass becomes identical to the rest mass when placed in the framework of classical physics where the collective effects are described only through the elastic constant $K$. Therefore, in the framework of the elastic mode theory where $v \geq 1$, the transition from a dynamical regime where the ratio $m_{vi}^l/m_B$ remains nearly constant (which we call **subshearic**) to the regime (which we call **supershearic**) where it increases according to a power law in $\lambda$ satisfying Eq. (9) and (13), is obtained when $\lambda = 1$, i.e. when

$$\left( \frac{m_{vi}^l}{m_B} \right)_{\lambda=1} = \frac{1}{H_N(v, 1)} = \frac{1}{H_N^1(v)} \qquad (C-8)$$

The transition value of the ratio $\left(m^l_{vi}/m_B\right)_{\lambda=1}$ thus depends on the state of the medium through the variable $v$ but depends little on $N$ as soon as this one is large enough (i.e. $N > 10$) as it can be seen on Fig. 21.

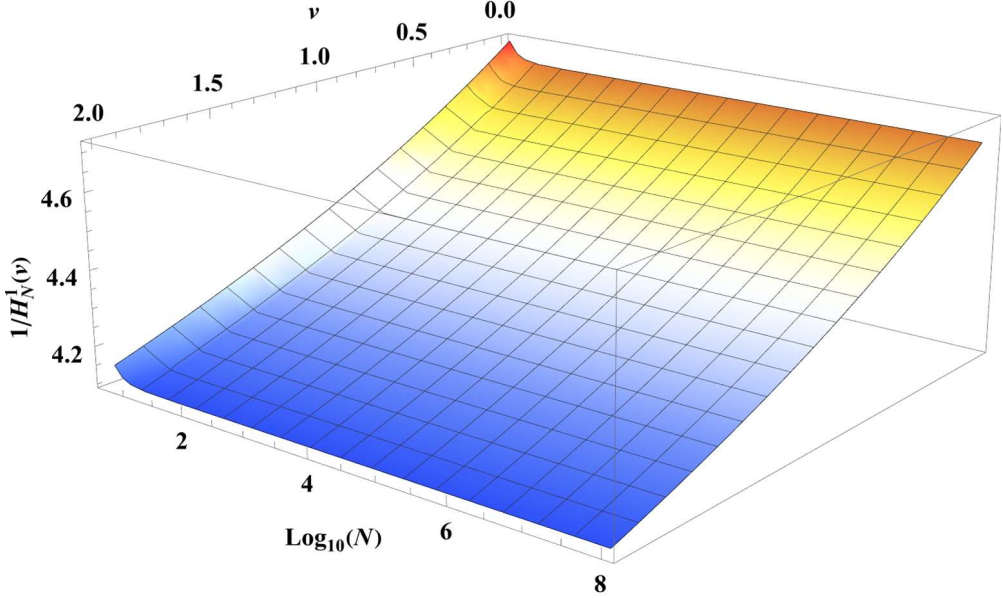

Fig. 21. Evolution of Eq. (C-8) versus $v$ and $N$. This surface separates the *subshearic* regime to the *supershearic* one.

We can remark that if again we consider classical physics, then Eq. (C-8) leads to the value $H^1_N(0)^{-1} = 4.65979...$ which is very close to the Feigenbaum number $\delta_F = 4.66920...$ (Ref. 20) which describes the self-similarity of an infinite cascade by period doubling in the framework of chaos theory (e.g. Ref. 21). As we mentioned in Ref. 6, the function $H_N$ is a self-similar function (e.g. at the transition $v = 1$, all derivatives of the function $H_N(v)$ are such that $\lim_{v \to 1} \dfrac{d^n}{dv^n} H_N(v) = \dfrac{(\ln N)^{n+1}}{n+1}$; so all derivatives are infinite for $N$ infinite) and one could expect to find the Feigenbaum number. The small difference can be explained by the fact that the cascade considered for example to describe the appearance of turbulence in Appendix B is never infinite but always finite. Under these conditions, we end up with a value that is always lower than the limit value for a quadratic logistic function (Ref. 20).

Fig. 22 shows the evolution of Eq. (C-2) for the value of $v$ corresponding to the experiments analyzed in this paper. It appears that for values of $\lambda$ lower than $10^{-10}$, $m^l_{vi}/m_B = 1/H_N(v,\lambda)$ tends to a constant smaller than 1. This region of $\lambda$ is the one typically corresponding to the Newtonian plateau. The subshearic regime where the viscous mass is smaller than the basic unit rest mass corresponds to the fact that in a laminar flow the dissipation is weak and therefore the equivalent in mass converted into heat is also small compared to the energy transported by the flow. According to Eq. (9), it can be observed with Fig. 22 that the supershearic regime no longer depends on $N$ as soon as $N$ is greater than 10.

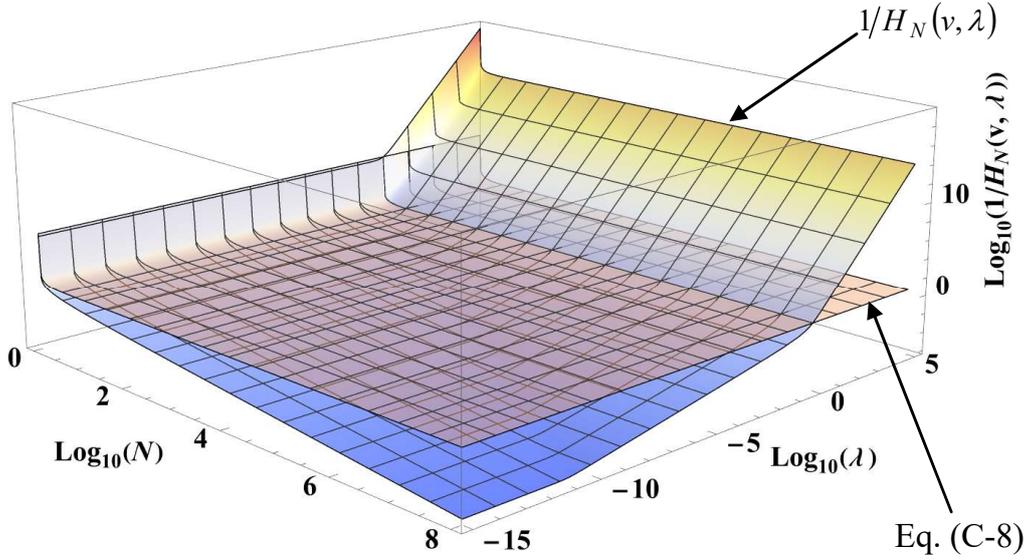

Fig. 22. Logarithmic plot of the function $1/H_N(v,\lambda)$ and Eq. (C-8) for $v = 1.71492$. The states below the semi-transparent light red surface correspond to the *subshearic* regime and above to the *supershearic* one.

As can be seen in Fig. 23, in the limiting case where $v = 2$ and $N$ is infinite then the function $H_N(v,\lambda)$ can be represented by a simple power law versus $\lambda$ in the subshearic regime. In other words, in this limit, outside a small region around $\lambda = 1$, the function $H_{N\to\infty}(v = 2,\lambda)$ can be approximated by two power laws as a function of $\lambda$ such that:

$$H_{N\to\infty}(v = 2, \lambda) \cong \begin{cases} \dfrac{1}{0.9\,\lambda^{3/4}} & \text{if } \lambda < 10^{-3} \\[2mm] \dfrac{1}{3\,\lambda^3} & \text{if } \lambda > 1 \end{cases} \qquad\text{(C-9)}$$

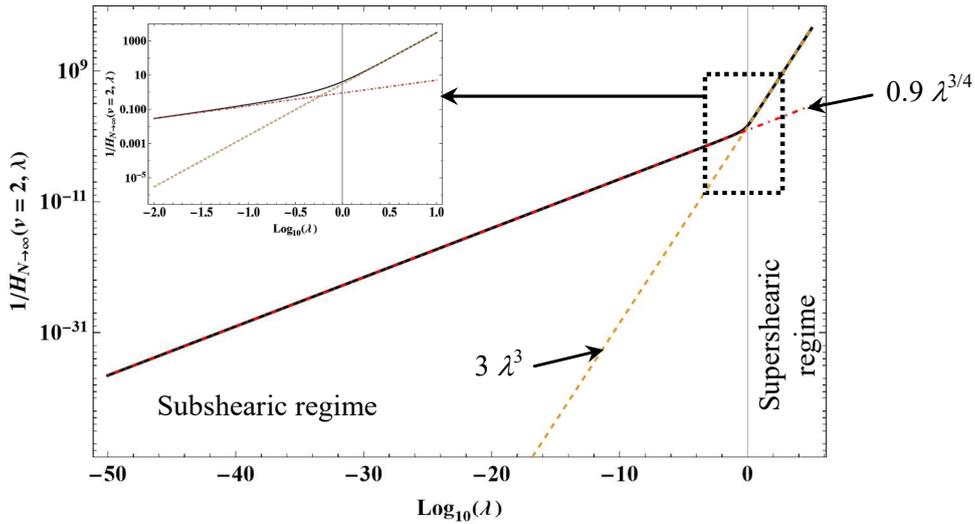

Fig. 23. Logarithmic plot of the function $1/H_N(v,\lambda)$ (black curve) and the inverse of Eq. (C-9) in the limiting case where $v = 2$ and $N \to \infty$. A zoom corresponding to the dotted black rectangle shows the transition zone between the two power laws.

In conclusion of this appendix, it appears that the present modeling is based on a new notion of mass which depends on the dynamic state of the system, contrary to classical physics where the notion of mass simply corresponds to the energy content of the matter in terms of weight. It is only in the framework of the SR that one can find similar ideas. However as mentioned above, the velocity $v_f$ in SR should not be confused with the average velocity between two particles $\overline{v}_m$ involved in the parameter $\lambda$. Moreover, the dependence of the mass on the dynamical state in SR is associated with the transport of energy from a source to another point in space, without any loss of energy between the source and the detector (i.e. the energy provided by the source is time independent), while in the present model, the situation is quite different. Indeed, in an ordinary laminar flow, the kinetic energy transported per unit volume is $\frac{1}{2}\rho\, v_f^2$, where $\rho$ is the density of the fluid calculated with the rest mass, but there is dissipation in the sense that part of this energy is lost as heat in this transport, this loss being characterized by the viscous dissipation time $\tau_v = \dfrac{d^2}{\eta/\rho} \approx \tau\, H_N(v)$. The generalization of the SR within the framework of the inertial mode theory will be described in a future paper.